\documentclass[%
 reprint,amsmath,amssymb, aps,prl]{revtex4-1}
\usepackage{graphicx}
\usepackage{grffile}
\usepackage{ulem}
\usepackage{dcolumn}
\usepackage{bm}
\usepackage{hyperref}

\usepackage{color}

\begin{document}
\preprint{ }
\title{Two-temperature activity induces liquid-crystal phases inaccessible in equilibrium}

\author{Jayeeta Chattopadhyay, Sriram Ramaswamy, Chandan Dasgupta and Prabal K. Maiti}

\email{maiti@iisc.ac.in}
\affiliation{%
Centre for Condensed Matter Theory, Department of Physics, Indian Institute of Science, Bangalore 560012, India
}%
\date{\today}



\begin{abstract}
In equilibrium hard-rod fluids, and in effective hard-rod descriptions of anisotropic soft-particle systems, the transition from the isotropic (I) phase to the nematic phase (N) is observed above the rod aspect ratio $ L/D = 3.70 $ as predicted by Onsager. We examine the fate of this criterion in a molecular dynamics study of a system of soft repulsive spherocylinders rendered active by coupling half the particles to a heat bath at a higher temperature than that imposed on the other half. 
We show that the system phase-separates and self-organizes into various liquid-crystalline phases that are not observed in equilibrium for the respective aspect ratios. In particular, we find a nematic phase for $L/D = 3$ and a smectic phase for $L/D = 2$ above a critical activity. 

\end{abstract}


          
\maketitle



The equilibrium liquid-crystalline properties of anisotropic particles are well understood \cite{onsager, de-gennes, bolhuis, McGrother, cuetos-2002, cuetos-2015, maiti2002,lansac2003,thirumalai1986effect}. Onsager \cite{onsager} first developed an analytical theory to describe the transition from the isotropic (I) phase to the nematic (N) phase, which has uniaxial apolar orientational order, and predicted that a phase with purely nematic order, breaking no other symmetries, cannot arise for hard rods with aspect ratio $L/D < 3.70$ \cite{onsager, bolhuis, de-gennes}. In this Letter, we inquire into the extension of Onsager's limit to active matter, in the specific context of two-temperature systems.

Active matter is driven locally by a constant supply of free energy to its constituent particles, which dissipate it by performing mechanical work \cite{Sriram1,Sriram2,toner2005,Romanczuk_2012,prost2015active,ramaswamy2019active,gompper20202020,PhysRevLett.89.058101,PhysRevX.12.010501,Yeomans-PNAS-2012,Yeomans-Natcomm-2018,Yeomans-PRL-2020,AnanyoPRL2020,Yeomans-PRL-2022,Hartmut2013,Hartmut2015,Hartmut-PRX-2015,Hartmut2016,Hartmut2019,Hartmut2020,Speck2014,Baskaran-softmatter,redner-baskaran}.
In flocking models, arguably the most familiar examples, activity is linked to a vector order parameter 
\cite{Sriram1,vicsek2012,Sriram2,toner-Tu-1995,toner2005}.
In scalar active matter\cite{wittkowski2014scalar,Golestanian-PRX-2020,joanny-2015,frey,Hartmut-PRX-2015}, activity enters by minimally breaking detailed balance in scalar Halperin-Hohenberg models \cite{RevModPhys.49.435} or, as in our present work, by introducing two (or more) species of particles coupled to thermal baths at distinct temperatures \cite{joanny-2015,joanny-2018,joanny-2020,frey,ganai,kremer1,kremer2,Siva,Active-SRS,https://doi.org/10.48550/arxiv.2109.00415}. The temperatures in question should be viewed not as thermodynamic but emergent, arising from the effective diffusivities of the multiple motile species. Two-temperature models have accounted for chromatin organization in the cell nucleus \cite{ganai}; phase-separation and self-organized structure formation in binary mixtures of Brownian soft disks \cite{frey} and Lennard-Jones (LJ) particles \cite{Siva}, and in polymer systems \cite{kremer1,kremer2}. In a recent work \cite{Active-SRS}, we have implemented this idea in a system of soft repulsive spherocylinders (SRSs) of aspect ratio $L/D = 5$ (where $L$ and $D$ are the effective length and diameter defined by the anisotropic repulsive potential \cite{allen1993hard,vega1994fast,bolhuis,cuetos-2002,earl2001}) and showed that increasing the temperature of the hot particles promotes liquid-crystal ordering in the cold particles, shifting the I-N phase boundary to lower densities than its equilibrium location. Here, we aim to explore ordering transitions of SRS of different $L/D$, in particular, those below Onsager's limit. 

\begin{figure} [!htb]
	\centering
	\includegraphics[scale=1.5]{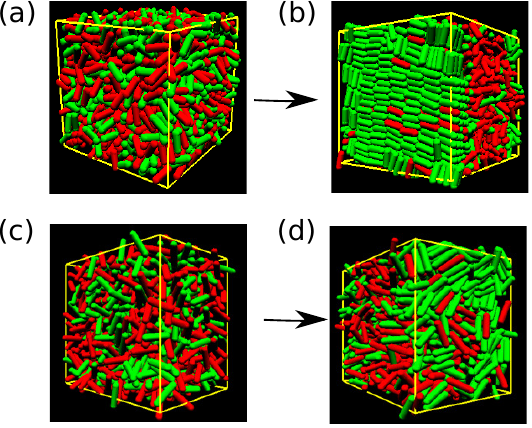}
	\caption{Snapshots representing steady state configurations of hot (red) and cold (green) particles before (left panel) and after (right panel) phase separation for the aspect ratios (a,b) $L/D = 2$ at the packing fraction $\eta = 0.45$ and activity $\chi = 9$; (c,d) $L/D = 3$ at $\eta = 0.33$ and $\chi = 4$. For each of the cases, the initial system is in homogeneous isotropic phase. After phase segregation, cold particles show (b) smectic ordering for $L/D = 2$ and (d) nematic ordering for $L/D = 3$ at the aforementioned activities. Hot particles remain in the isotropic phase with reduced density.} \label{phase-sep-diffL}

\end{figure}



To do so, we have carried out a series of Molecular Dynamics (MD) simulations of a system of SRS with $L/D = 5,3$ and $2$. The model and simulation protocol are the same as those in our previous work \cite{Active-SRS}. Here we give brief details for completeness. The SRS interact through the Weeks-Chandler-Andersen potential \cite{weeks1971}: 
\begin{equation} 
\begin{split}
U_{SRS} & =  4\varepsilon\left[\left(\dfrac{D}{d_{m}}\right)^{12}-\left(\dfrac{D}{d_{m}}\right)^{6}\right]+\epsilon \qquad \hbox{if} \quad d_{m} < 2^{\frac{1}{6}}D\\
& = 0 \qquad \qquad \qquad \qquad \qquad \qquad \qquad \hbox{if} \quad d_{m} \geq 2^{\frac{1}{6}}D .
\end{split}
\label{wca-eq}
\end{equation} 
Here $d_{m}$, the shortest distance between two spherocylinders, implicitly determines their relative orientation and the direction of the interaction force \cite{allen1993hard,vega1994fast,bolhuis,cuetos-2002,earl2001}. We build the initial configuration in a hexagonal close-packed (HCP) crystal structure and perform MD simulations at constant particle number, pressure and temperature (NPT) with periodic boundary conditions in all three directions \cite{rotunno2004,maiti2002,lansac2003}. For each aspect ratio, we simulate a wide range of pressures spanning the  transition from the crystal to the isotropic phase and characterize the phases by calculating the nematic order parameter and appropriate pair correlation functions. For a system of $N$ spherocylinders labelled $i = 1, \ldots , N$, with orientations defined by unit vectors ${\bf u}_{i}$ with components $u_{i\alpha}$, the traceless symmetric nematic order parameter $\bf Q$ has components $Q_{\alpha\beta} = ({1}/{N})\sum_{i=1}^{N}\left[({3}/{2})u_{i\alpha}u_{i\beta}-\dfrac{1}{2}\delta_{\alpha\beta}\right]$.
The scalar nematic order parameter $S$ is the largest eigenvalue of $\bf Q$. \\

Hereafter we work in reduced units defined in terms of the system parameters $\epsilon$ and $D$: temperature $T^{*}=k_{B}T/\epsilon $, pressure $ P^{*}= Pv_{hsc}/(k_{B}T)$, packing fraction $ \eta = v_{hsc}\rho$, where $ \rho = N/V $ and $ v_{hsc}= \pi D^{2}(D/6 + L/4)$ is the volume of a spherocylinder. 

Activity in our system is introduced by connecting half of the particles to a thermostat of higher temperature, while maintaining the temperature of the other half fixed at a lower value. 
Let $T_{h}^{*}$ and $T_{c}^{*}$ be the temperatures of the baths connected to the hot and cold particles respectively, controlled by a Berendsen thermostat \cite{berendsen1984} with a time constant $\tau_{T}=0.01$. We then define the activity $\chi = (T_{h}^{*} -  T_{c}^{*})/T_{c}^{*}$. Starting from a statistically isotropic structure at a definite temperature with $T_{h}^{*} = T_{c}^{*} = 5$, we gradually increase the temperature of the hot particles $T_{h}^{*}$, keeping the \textit{volume} of the simulation box constant (NVT ensemble) \cite{Active-SRS}.



In equilibrium, we observe four stable phases for $L/D = 5$: crystal (K), smectic-A (SmA), nematic (N), isotropic (I); three stable phases for $L/D = 3$: crystal, smectic-A, isotropic and two stable phases for $L/D = 2$: crystal, isotropic (Fig. S1 in Supplemental Information (SI) \cite{sp}). Our results are consistent with those of a previous study carried out by Cuetos et al. \cite{cuetos-2002,cuetos-2015} for soft rods. 

To define a criterion analogous to that of Onsager \cite{onsager} for our case, we construct an effective hard-cylinder diameter for the SRS, in terms of the interaction potential and the temperature of the system [Fig. S2], as in Cuetos {\it et al.} \cite{cuetos2005parsons,cuetos-2015, boublik1976}:
\begin{equation}
D_{eff}(T) = \int_{0}^{\infty} (1-\exp[-\beta U_{SRS}(d_{m})] dd_{m}  . \label{eq-Deff} 
\end{equation}
Therefore, a SRS with aspect ratio  $A_{SRS} = L/D $ can be mapped to a HSC with an effective aspect ratio $A_{HSC} = L/D_{eff}$.
In Table \ref{tab-2}, we mention different values of $A_{SRS}$ and the corresponding values of $A_{HSC}$ at $ T^{*} = 5$.
Using Eq. \ref{eq-Deff}, the value of $A_{SRS}$ corresponding to $A_{HSC} = 3.70$ (Onsager's limit for HSC) becomes $A_{SRS} = 3.52$ (Onsager's limit for SRS) at $ T^{*} = 5$. This approximate version of Onsager's criterion is verified by the absence of a nematic phase for SRS at thermal equilibrium with $A_{SRS}=3$ and $2$ in our simulations and those of Cuetos {\it et al.} \cite{cuetos-2002,cuetos-2015}.

\begin{figure*} [!htb]
	\centering
	\includegraphics[scale=0.9]{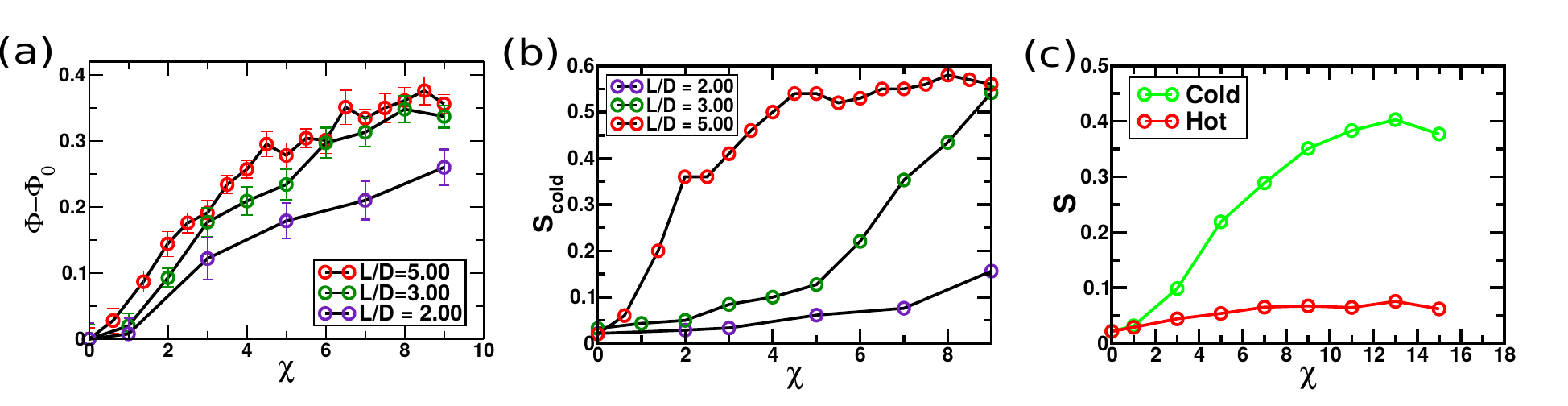}
	\caption{ (a) Density order parameter $\phi$ of the system and (b) nematic order parameter of the cold particles $S_{cold}$ vs activity $ \chi $ for different aspect ratios $L/D$ at their respective packing fractions: for $L/D = 5$, $\eta = 0.36$ ; for $L/D = 3$, $\eta = 0.33$  and for $L/D = 2$, $\eta = 0.35$. (c) Nematic order parameter of the cold and hot particles for $L/D = 2$ at a higher packing fraction $\eta = 0.45 $. $\phi_{0}$ designates the magnitude of $\phi$ in initial system (at $\chi = 0$).} \label{denop-s}
	
\end{figure*}

\begin{figure*} [!htb]
	\centering
	\includegraphics[scale=0.8]{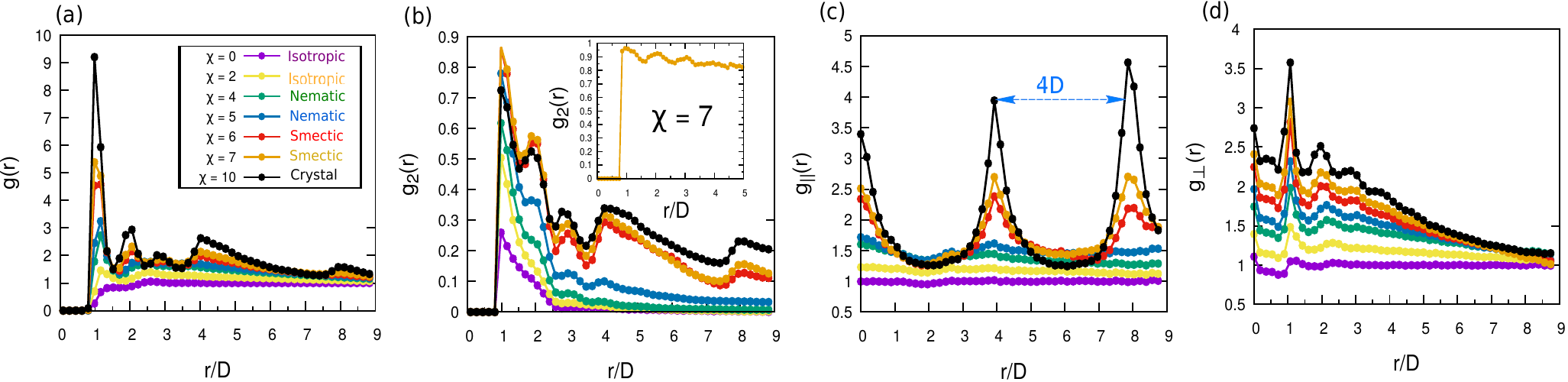}
	\caption{ Pair correlation functions for the cold particles at different activities $\chi$ for $L/D = 3$ at the packing fraction $\eta = 0.33$. (a) The center of mass pair radial distribution function $g(r)$, (b) orientational pair distribution function $g_{2}(r)$, (c) projection of $g(r)$ along the direction parallel $[g_{\parallel} (r)]$ and (d) perpendicular $[g_{\perp} (r)]$ to the director of the spherocylinders. 
	The inset of Fig. (b) shows $g_{2}(r)$ at $\chi = 7$ in a single cluster with a definite director. The distance between the two peaks in panel (c) is 4D which is the end to end distance of a spherocylinder with $L/D = 3$.} \label{g-L3}

\end{figure*}

Starting from a homogeneous isotropic structure, we observe local phase separation between hot and cold particles, which emerges at the macroscopic scale by forming a well-defined interface (Fig. \ref{phase-sep-diffL}). The phase separation is quantified through the difference between the local densities of hot and cold particles. Dividing the simulation box into sub-cells labelled $i = 1, ..., N_{cell}$ and letting $n^{i}_{hot}$, $n^{i}_{cold}$ be the numbers of hot and cold particles respectively in each cell, we define 
$\phi=N_{cell}^{-1} \langle\sum_{i=1}^{N_{cell}} {\lvert(n^{i}_{hot}-n^{i}_{cold})\rvert}/{n^{i}_{tot}}\rangle$ \cite{Siva, Active-SRS} where the average $\langle ... \rangle$ is carried out over a sufficiently large number of steady-state configurations. The choice of $N_{cell}$ depends on the aspect ratio of the rod. It is chosen such that each cell contains a sufficient number of particles to get stable statistics.

In Fig. \ref{denop-s}-(a), we plot $\phi$ as a function of activity $\chi$ for $L/D = 5, 3$ and $2$. For the sake of comparison, the system is chosen with a packing fraction between $\eta = 0.33-0.36$ for which the system is in the isotropic phase at thermal equilibrium for the given $L/D$ at $T^* = 5$ (Fig. S1).
From Fig. \ref{denop-s}-(a), we observe: (i) phase separation starts at a lower activity for higher aspect ratios; (ii) the amount of phase separation at a given $\chi$ is higher for higher aspect ratios. 
To understand observation (i) precisely, we calculate the critical activity $\chi_{c}$, which is defined as the value of $\chi$ above which macroscopic phase separation is seen (see SI and Fig. S3 for details \cite{sp}). The calculated ranges of critical activities for the given packing fractions are:  $\chi_{c} = 1.4-2 $ for $L/D = 5$, $\chi_{c} = 2-3$ for $L/D = 3$ and $\chi_{c} = 3-5$ for $L/D = 2$. 

After phase separation, the interactions between hot and cold particles mostly take place at the interface. The cold zone undergoes an ordering transition above a critical activity $\chi^{*}$ that depends on the aspect ratio of the rods as well as their packing fractions.
In Fig. \ref{denop-s}-(b), we see that the nematic order parameter of cold particles $S_{cold}$ increases with $\chi$ for $L/D = 5$ and $3$ for $0 \le \chi \le 9 $, while for $L/D = 2 $ it increases above $\chi \ge 10$ (Fig. S4) resulting in a higher $\chi^{*}$ .
However, the difference between the critical activities for phase separation and ordering decreases with increasing density. In Fig \ref{denop-s}-(c), we have plotted $S_{cold}$ versus $\chi$ for $L/D = 2$ at a slightly higher packing fraction, $\eta = 0.45$, which also corresponds to the isotropic phase in equilibrium. Here we see both phase separation and ordering transition for the given range of activities. The critical activities for phase separation and ordering at $\eta = 0.45$ are $\chi_{c} = 3-5$ and $\chi^{*} = 5$, respectively.\\


The phases of the ordered structures in the cold zones are identified by calculating the local nematic order parameter $S$ and appropriate pair correlation functions (see SI for details). We observe that with increasing activity both translational and orientational correlations in the cold zone are enhanced, indicating incipient order in the cold zone. Interestingly, we observe liquid crystal phases for small values of $L/D$ that do not occur  for the same parameter range in a one-temperature system, i.e., in equilibrium.

In Fig. \ref{g-L3} (a-d), we plot different pair correlation functions of the cold particles at different activities for $L/D = 3$ at $\eta =0.33$. We see that, both of translational $[g(r)]$ and orientational $[g_{2}(r)]$ pair correlation functions are flat in absence of activity ($\chi = 0$) which is obvious for an isotropic phase. $g(r)$ develops the first peak at $\chi = 4$ and eventually the other peaks at higher values of $\chi$. In Fig. \ref{g-L3}(b), we see that, $g_{2}(r)$ has a finite correlation length of roughly $2.5D$, beyond which it decays to zero for $\chi = 4,5 $ and to a finite value for $\chi \ge 6$. 
This is also observe in the calculation of the half width at half maximum (HWHM) of $g_{2}(r)$ defined as the distance from the first peak at which the value of $g_2(r)$ is half of its value at the first peak [see Fig. S5 in SI for details].
These observations suggest that there exists a finite orientational order in the cold zone for $\chi = 4, 5$ which designates this phase as nematic.
Above this activity, $g_{2}(r)$ develops multiple peaks at longer distances and saturates at a finite value. This is due to the presence of multiple clusters of different average directors, which effectively suppress the overall orientational correlation. In these cases, we calculate the $g_{2}(r)$ in a single cluster of a definite director and find it to saturate at a higher value, as shown in the inset of Fig. \ref{g-L3}(b). Smectic and crystalline structures are identified by calculating translational correlations along the parallel [$g_{\parallel}(r)$] and perpendicular [$g_{\perp}(r)$] directions of the average nematic director of the spherocylinders. The periodic oscillations in $g_{\parallel}(r)$ and liquid-like structure in $g_{\perp}(r)$ at $\chi = 6, 7$ indicate that the phase is smectic, as shown in Fig. \ref{g-L3} (c, d). 

Similarly, we find that the system with $L/D = 2 $ exhibits smectic ordering at $\eta = 0.45$, $\chi = 9 $ as shown in Fig. S6. 
However, the hot zone shows isotropic structure with reduced packing fractions for each of the cases. The snapshots of the different liquid crystal phases for the respective $L/D$s are shown in Fig. \ref{phase-sep-diffL}. We summarize these results in Table \ref{tab-2} where we mention the possible phases in  equilibrium and the emergent phases in active systems for the respective aspect ratios. 


\begin{table}[ht]
	\caption{Liquid crystal phases in the cold zone for the respective aspect ratios at different activities. The phases that are absent in the equilibrium system and occur in the active systems are mentioned in bold. The equilibrium phases of HSC at the given aspect ratios are taken from Ref. \cite{bolhuis,McGrother} : }
	\vspace{0.4cm}
	\centering
	\begin{tabular}{|c|c|c|c|}
		\hline
		
		$ A_{SRS}$ & $ A_{HSC} $ & Equilibrium Phases & Phases at $\chi \neq 0 $ \\
		at $T^{*} = 5$ & & at $\chi = 0 $ & \\[3ex]
		\hline
		
		5 & 5.28 & I, N, Sm, K & I, N, Sm, K, \\
		 & & & Multi-domain K \\  [1.5ex]
		
		3 & 3.20 &I, Sm, K & I, \textbf{N}, Sm, K \\ [1.5ex]
		
		2 & 2.11  & I, K & I, \textbf{Sm}, K \\ [1.5ex]
		
		\hline

	\end{tabular}
	
	\label{tab-2}
\end{table}

\begin{figure*} [!htb]
	\centering
	\includegraphics[scale=0.75]{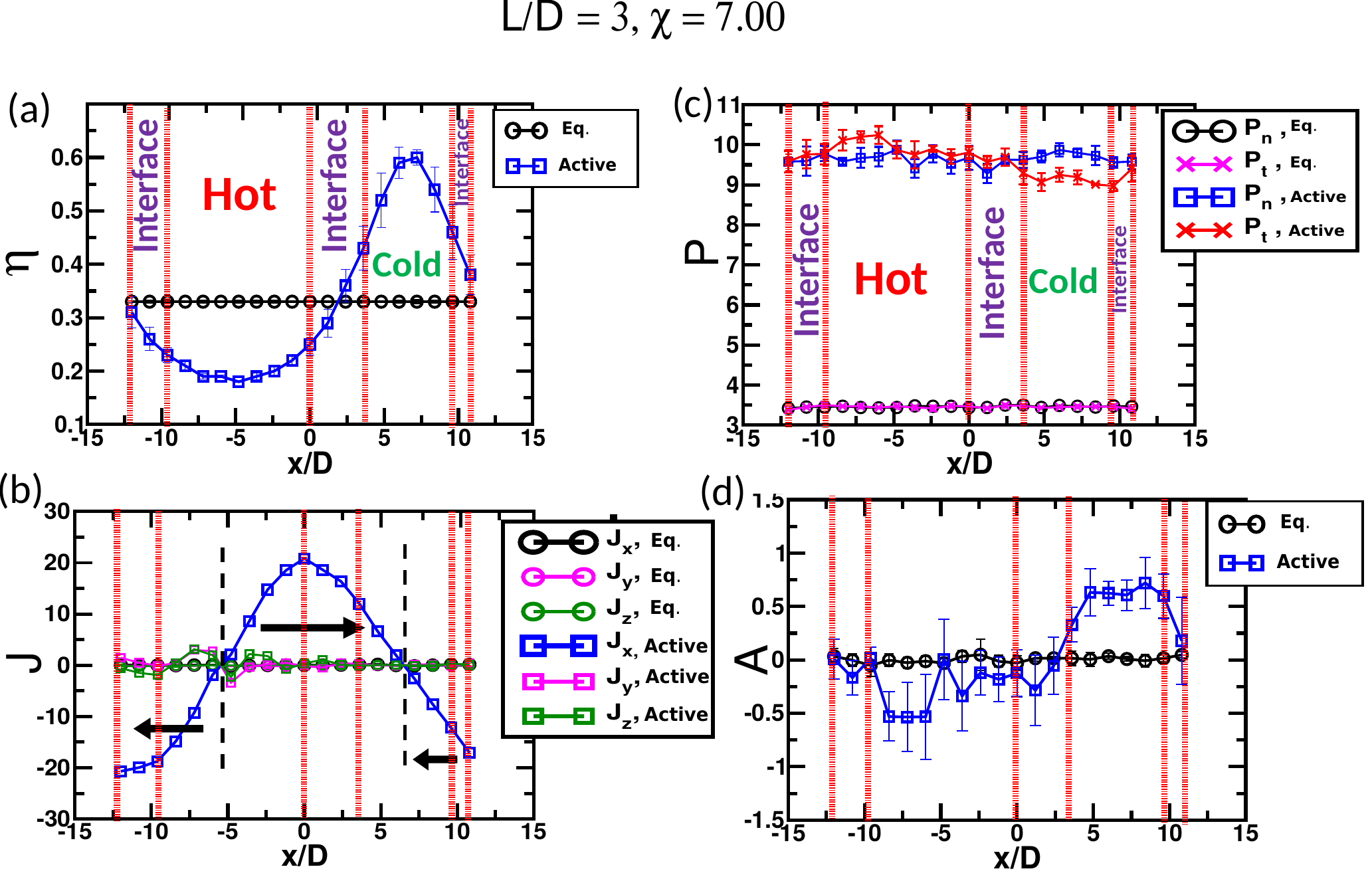}
	\caption{Pressure anisotropy and local heat flux across the hot-cold interface for $L/D = 3$ at $\eta = 0.33 $ and $\chi = 7$. We plot (a) effective packing fraction $\eta$; (b) local heat flux $J$; (c) normal ($P_{n}$) and tangential ($P_{t}$) components of the pressure and (d) pressure anisotropy $A = P_{n}-P_{t}$ along the direction perpendicular to the interfacial plane. The red dotted lines indicate boundaries of different zones. 
	The black dashed lines in Fig. (b) indicate the location where the heat flux becomes $0$ . The arrows indicate the directions of the heat flow depending on the sign of $J$.
	Here, we find a positive pressure anisotropy at the interface that extends in the cold zone and a finite heat flux flowing from the bulk hot zone to the bulk cold zone.} \label{stress-ana}
	\end{figure*}
	

To find the microscopic origin of the ordering transitions that are not observed in equilibrium, we calculate the local pressure in the phase-separated system. We divide the simulation box into a number of slabs ($i$) along the direction normal to the interface, following the procedure mentioned in Ref. \cite{Active-SRS}. The location of the interface is identified by calculating the local density of the hot and cold particles. The region where the local densities change sharply between their values in the segregated zones is identified as the interfacial region [Fig. \ref{stress-ana}(a)].

We then calculate the pressure components along the normal and tangential directions of the interfacial plane using the diagonal components of the stress tensor. We designate the normal direction of the interface as $x$ and the two tangential directions of the interfacial plane as $y$ and $z$. Therefore, the normal $ P_{n}$ and tangential $ P_{t}$ components of the pressure are defined as: $P_{n}(i)= P_{xx}(i)$ and $P_{t}(i)= (P_{yy}(i) + P_{zz}(i))/2 $, 
where,
\begin{equation}
    P_{\alpha\beta}(i) = \frac{1}{V(i)}\left(\sum_{j=1}^{n(i)}  mv_{j}^{\alpha}v_{j}^{\beta}+\sum_{j=1}^{n(i)-1} \sum_{k > j}^{}r_{jk}^{\alpha}f_{jk}^{\beta}\right).
    \label{press_eq}
\end{equation}

The $1^{st}$ and $2^{nd}$ term in Eq. \ref{press_eq} represent the kinetic and virial contributions (arises due to the particle's interaction) of the pressure tensor, where, 
$\vec{r_{j}} = (r_{j}^{\alpha})$ and
$\vec{v_{j}} = (v_{j}^{\alpha})$ are
the position and velocity of the
$j^{th}$ particle $(\alpha = x, y,
z)$ and $\vec{r_{jk}}$ and
$\vec{f_{jk}}$ are the relative distance and interacting force between the SRSs $j$ and $k$. $n(i)$ and $V(i)$ represent total number of particles and volume of the $i^{th}$ slab.

In Figs. \ref{stress-ana}(c), for the aspect ratio $L/D = 3 $, we find that, (i) in equilibrium ($\chi = 0 $), normal and tangential pressure components are equal, implying that the pressure tensor is isotropic throughout the simulation box. (ii) After phase separation ($\chi \geq \chi_{c}$), $P_{n}$ and $P_{t}$ are equal (within the error bars) in the hot zone. Thus, the pressure anisotropy is roughly equal to zero in this region.
(iii) At the interface, the tangential pressure decreases and acquire a lower value in the cold zone. However, the normal pressure remains balanced throughout the simulation box. 
This causes a pressure anisotropy at the hot-cold interface that persists in the bulk cold zone as well [Fig.\ref{stress-ana}(d)]. 
The pressure anisotropy increases with activity (see Fig. S7) and acquires a finite positive value in the cold zone.
This causes an effective compression of the cold zone along the normal direction of the interface that may drive the cold 
particles to orient along the parallel direction of the interfacial plane, thereby inducing an ordering transition.

To understand further the effects of the activity, we calculate the local heat flux $\vec{J}$ in each slab using the following equation \cite{hansen2013theory,Pranab-PRE}
\begin{equation}
    \vec{J}(i) = \dfrac{1}{V(i)}\langle\sum_{j = 1}^{n(i)} \vec{v_{j}}e_{j} + \sum_{j = 1}^{n(i)} \mathbf{\sigma_{j}} \cdot \vec{v_{j}} \rangle ,
\end{equation}
where $e_{j} = ({1}/{2})mv_j^{2} + \sum_{j \neq k} U_{jk}$ is the total energy and $\mathbf{\sigma}_{j} = ({1}/{2}) \sum_{j \neq k} \vec{r_{jk}}  \vec{f_{jk}}$ is the stress tensor. 
In Fig. \ref{stress-ana}(b), we show the spatial variation of $\vec{J}$ along the direction normal to the interface. We find $\vec{J} = 0$ in equilibrium, but it obtains a finite value in the phase separated systems, both at the interface and in the bulk region. 
This reveals that, though the interaction between hot and cold particles takes place mainly at the interface, its effect extends to the bulk region as well. The sign of $\vec{J}$ indicates heat flows from the bulk hot to the bulk cold zone. This results in heterogeneous activity and broken time-reversal invariance throughout the simulation box giving rise to anomalous thermodynamic behavior (such as pressure anisotropy) away from the interface. \\

To check for possible finite size effects, we have done a similar analysis for larger system sizes with $N = 4608$ for $L/D = 2 $ and $N = 3456 $ for $L/D = 3$. We observe a similar trend of ordering transition for the respective aspect ratios. Here also we observe a nematic phase for $L/D =3$ and a smectic phase for $L/D=2$ (See SI for details, Fig. S8-S11). Further increase of activity causes the cold zone to crystallize.\\



Our simulation study examines the effect of two-temperature activity in a soft-rod fluid for a range of effective aspect ratios of the rods. 
We show that the two-temperature model can give rise to liquid crystal phases that are not observed in equilibrium for the respective aspect ratios. We observe a smectic phase for $L/D = 2$ and a nematic phase for $L/D = 3$. We find that the presence of two temperatures causes a pressure anisotropy extending from the hot-cold interface into the bulk of the cold zone, and a heat current flowing from the hot to the cold zone. Thus, the nonequilibrium behavior is not limited to the hot-cold interfaces  but pervades the system as a whole, driving the anomalous ordering transitions in the cold zone. An understanding of these results within analytical theory, experimental realizations of two-temperature systems, presumably in suspensions with bidisperse motility, and methods to capture and stabilize the anomalously ordered domains are some of the challenges that emerge from our work.



We would like to thank Prof. Aparna Baskaran for helpful discussions. We thank SERB, India for financial support through providing computational facility. JC acknowledges support through an INSPIRE fellowship. JC thanks Tarun Maity, Subhadeep Dasgupta and Swapnil Kole for insightful discussions. SR was supported by a J C Bose Fellowship of the SERB, India, and by the Tata Education and Development Trust, and acknowledges discussions during the KITP 2020 online program on Symmetry, Thermodynamics and Topology in Active Matter. This research was supported in part by the National Science Foundation under Grant No. NSF PHY-1748958. CD was supported by a Distinguished Fellowship of the SERB, India.


\bibliography{ref}

\providecommand{\noopsort}[1]{}\providecommand{\singleletter}[1]{#1}%
\begin{thebibliography}{58}%
\makeatletter
\providecommand \@ifxundefined [1]{%
 \@ifx{#1\undefined}
}%
\providecommand \@ifnum [1]{%
 \ifnum #1\expandafter \@firstoftwo
 \else \expandafter \@secondoftwo
 \fi
}%
\providecommand \@ifx [1]{%
 \ifx #1\expandafter \@firstoftwo
 \else \expandafter \@secondoftwo
 \fi
}%
\providecommand \natexlab [1]{#1}%
\providecommand \enquote  [1]{``#1''}%
\providecommand \bibnamefont  [1]{#1}%
\providecommand \bibfnamefont [1]{#1}%
\providecommand \citenamefont [1]{#1}%
\providecommand \href@noop [0]{\@secondoftwo}%
\providecommand \href [0]{\begingroup \@sanitize@url \@href}%
\providecommand \@href[1]{\@@startlink{#1}\@@href}%
\providecommand \@@href[1]{\endgroup#1\@@endlink}%
\providecommand \@sanitize@url [0]{\catcode `\\12\catcode `\$12\catcode
  `\&12\catcode `\#12\catcode `\^12\catcode `\_12\catcode `\%12\relax}%
\providecommand \@@startlink[1]{}%
\providecommand \@@endlink[0]{}%
\providecommand \url  [0]{\begingroup\@sanitize@url \@url }%
\providecommand \@url [1]{\endgroup\@href {#1}{\urlprefix }}%
\providecommand \urlprefix  [0]{URL }%
\providecommand \Eprint [0]{\href }%
\providecommand \doibase [0]{http://dx.doi.org/}%
\providecommand \selectlanguage [0]{\@gobble}%
\providecommand \bibinfo  [0]{\@secondoftwo}%
\providecommand \bibfield  [0]{\@secondoftwo}%
\providecommand \translation [1]{[#1]}%
\providecommand \BibitemOpen [0]{}%
\providecommand \bibitemStop [0]{}%
\providecommand \bibitemNoStop [0]{.\EOS\space}%
\providecommand \EOS [0]{\spacefactor3000\relax}%
\providecommand \BibitemShut  [1]{\csname bibitem#1\endcsname}%
\let\auto@bib@innerbib\@empty
\bibitem [{\citenamefont {Onsager}(1949)}]{onsager}%
  \BibitemOpen
  \bibfield  {author} {\bibinfo {author} {\bibfnamefont {L.}~\bibnamefont
  {Onsager}},\ }\href@noop {} {\bibfield  {journal} {\bibinfo  {journal}
  {Annals of the New York Academy of Sciences}\ }\textbf {\bibinfo {volume}
  {51}},\ \bibinfo {pages} {627} (\bibinfo {year} {1949})}\BibitemShut
  {NoStop}%
\bibitem [{\citenamefont {De~Gennes}\ and\ \citenamefont
  {Prost}(1993)}]{de-gennes}%
  \BibitemOpen
  \bibfield  {author} {\bibinfo {author} {\bibfnamefont {P.-G.}\ \bibnamefont
  {De~Gennes}}\ and\ \bibinfo {author} {\bibfnamefont {J.}~\bibnamefont
  {Prost}},\ }\href@noop {} {\emph {\bibinfo {title} {The physics of liquid
  crystals}}},\ Vol.~\bibinfo {volume} {83}\ (\bibinfo  {publisher} {Oxford
  university press},\ \bibinfo {year} {1993})\BibitemShut {NoStop}%
\bibitem [{\citenamefont {Bolhuis}\ and\ \citenamefont
  {Frenkel}(1997)}]{bolhuis}%
  \BibitemOpen
  \bibfield  {author} {\bibinfo {author} {\bibfnamefont {P.}~\bibnamefont
  {Bolhuis}}\ and\ \bibinfo {author} {\bibfnamefont {D.}~\bibnamefont
  {Frenkel}},\ }\href@noop {} {\bibfield  {journal} {\bibinfo  {journal} {The
  Journal of chemical physics}\ }\textbf {\bibinfo {volume} {106}},\ \bibinfo
  {pages} {666} (\bibinfo {year} {1997})}\BibitemShut {NoStop}%
\bibitem [{\citenamefont {McGrother}\ \emph {et~al.}(1996)\citenamefont
  {McGrother}, \citenamefont {Williamson},\ and\ \citenamefont
  {Jackson}}]{McGrother}%
  \BibitemOpen
  \bibfield  {author} {\bibinfo {author} {\bibfnamefont {S.~C.}\ \bibnamefont
  {McGrother}}, \bibinfo {author} {\bibfnamefont {D.~C.}\ \bibnamefont
  {Williamson}}, \ and\ \bibinfo {author} {\bibfnamefont {G.}~\bibnamefont
  {Jackson}},\ }\href {\doibase 10.1063/1.471343} {\bibfield  {journal}
  {\bibinfo  {journal} {The Journal of Chemical Physics}\ }\textbf {\bibinfo
  {volume} {104}},\ \bibinfo {pages} {6755} (\bibinfo {year}
  {1996})}\BibitemShut {NoStop}%
\bibitem [{\citenamefont {Cuetos}\ \emph {et~al.}(2002)\citenamefont {Cuetos},
  \citenamefont {Mart{\i}nez-Haya}, \citenamefont {Rull},\ and\ \citenamefont
  {Lago}}]{cuetos-2002}%
  \BibitemOpen
  \bibfield  {author} {\bibinfo {author} {\bibfnamefont {A.}~\bibnamefont
  {Cuetos}}, \bibinfo {author} {\bibfnamefont {B.}~\bibnamefont
  {Mart{\i}nez-Haya}}, \bibinfo {author} {\bibfnamefont {L.}~\bibnamefont
  {Rull}}, \ and\ \bibinfo {author} {\bibfnamefont {S.}~\bibnamefont {Lago}},\
  }\href@noop {} {\bibfield  {journal} {\bibinfo  {journal} {The Journal of
  chemical physics}\ }\textbf {\bibinfo {volume} {117}},\ \bibinfo {pages}
  {2934} (\bibinfo {year} {2002})}\BibitemShut {NoStop}%
\bibitem [{\citenamefont {Cuetos}\ and\ \citenamefont
  {Mart{\'\i}nez-Haya}(2015)}]{cuetos-2015}%
  \BibitemOpen
  \bibfield  {author} {\bibinfo {author} {\bibfnamefont {A.}~\bibnamefont
  {Cuetos}}\ and\ \bibinfo {author} {\bibfnamefont {B.}~\bibnamefont
  {Mart{\'\i}nez-Haya}},\ }\href@noop {} {\bibfield  {journal} {\bibinfo
  {journal} {Molecular Physics}\ }\textbf {\bibinfo {volume} {113}},\ \bibinfo
  {pages} {1137} (\bibinfo {year} {2015})}\BibitemShut {NoStop}%
\bibitem [{\citenamefont {Maiti}\ \emph {et~al.}(2002)\citenamefont {Maiti},
  \citenamefont {Lansac}, \citenamefont {Glaser},\ and\ \citenamefont
  {Clark}}]{maiti2002}%
  \BibitemOpen
  \bibfield  {author} {\bibinfo {author} {\bibfnamefont {P.~K.}\ \bibnamefont
  {Maiti}}, \bibinfo {author} {\bibfnamefont {Y.}~\bibnamefont {Lansac}},
  \bibinfo {author} {\bibfnamefont {M.~A.}\ \bibnamefont {Glaser}}, \ and\
  \bibinfo {author} {\bibfnamefont {N.~A.}\ \bibnamefont {Clark}},\ }\href@noop
  {} {\bibfield  {journal} {\bibinfo  {journal} {Physical review letters}\
  }\textbf {\bibinfo {volume} {88}},\ \bibinfo {pages} {065504} (\bibinfo
  {year} {2002})}\BibitemShut {NoStop}%
\bibitem [{\citenamefont {Lansac}\ \emph {et~al.}(2003)\citenamefont {Lansac},
  \citenamefont {Maiti}, \citenamefont {Clark},\ and\ \citenamefont
  {Glaser}}]{lansac2003}%
  \BibitemOpen
  \bibfield  {author} {\bibinfo {author} {\bibfnamefont {Y.}~\bibnamefont
  {Lansac}}, \bibinfo {author} {\bibfnamefont {P.~K.}\ \bibnamefont {Maiti}},
  \bibinfo {author} {\bibfnamefont {N.~A.}\ \bibnamefont {Clark}}, \ and\
  \bibinfo {author} {\bibfnamefont {M.~A.}\ \bibnamefont {Glaser}},\
  }\href@noop {} {\bibfield  {journal} {\bibinfo  {journal} {Physical Review
  E}\ }\textbf {\bibinfo {volume} {67}},\ \bibinfo {pages} {011703} (\bibinfo
  {year} {2003})}\BibitemShut {NoStop}%
\bibitem [{\citenamefont {Thirumalai}(1986)}]{thirumalai1986effect}%
  \BibitemOpen
  \bibfield  {author} {\bibinfo {author} {\bibfnamefont {D.}~\bibnamefont
  {Thirumalai}},\ }\href@noop {} {\bibfield  {journal} {\bibinfo  {journal}
  {The Journal of chemical physics}\ }\textbf {\bibinfo {volume} {84}},\
  \bibinfo {pages} {5869} (\bibinfo {year} {1986})}\BibitemShut {NoStop}%
\bibitem [{\citenamefont {Marchetti}\ \emph {et~al.}(2013)\citenamefont
  {Marchetti}, \citenamefont {Joanny}, \citenamefont {Ramaswamy}, \citenamefont
  {Liverpool}, \citenamefont {Prost}, \citenamefont {Rao},\ and\ \citenamefont
  {Simha}}]{Sriram1}%
  \BibitemOpen
  \bibfield  {author} {\bibinfo {author} {\bibfnamefont {M.~C.}\ \bibnamefont
  {Marchetti}}, \bibinfo {author} {\bibfnamefont {J.~F.}\ \bibnamefont
  {Joanny}}, \bibinfo {author} {\bibfnamefont {S.}~\bibnamefont {Ramaswamy}},
  \bibinfo {author} {\bibfnamefont {T.~B.}\ \bibnamefont {Liverpool}}, \bibinfo
  {author} {\bibfnamefont {J.}~\bibnamefont {Prost}}, \bibinfo {author}
  {\bibfnamefont {M.}~\bibnamefont {Rao}}, \ and\ \bibinfo {author}
  {\bibfnamefont {R.~A.}\ \bibnamefont {Simha}},\ }\href {\doibase
  10.1103/RevModPhys.85.1143} {\bibfield  {journal} {\bibinfo  {journal} {Rev.
  Mod. Phys.}\ }\textbf {\bibinfo {volume} {85}},\ \bibinfo {pages} {1143}
  (\bibinfo {year} {2013})}\BibitemShut {NoStop}%
\bibitem [{\citenamefont {Ramaswamy}(2010)}]{Sriram2}%
  \BibitemOpen
  \bibfield  {author} {\bibinfo {author} {\bibfnamefont {S.}~\bibnamefont
  {Ramaswamy}},\ }\href {\doibase 10.1146/annurev-conmatphys-070909-104101}
  {\bibfield  {journal} {\bibinfo  {journal} {Annual Review of Condensed Matter
  Physics}\ }\textbf {\bibinfo {volume} {1}},\ \bibinfo {pages} {323–345}
  (\bibinfo {year} {2010})}\BibitemShut {NoStop}%
\bibitem [{\citenamefont {Toner}\ \emph {et~al.}(2005)\citenamefont {Toner},
  \citenamefont {Tu},\ and\ \citenamefont {Ramaswamy}}]{toner2005}%
  \BibitemOpen
  \bibfield  {author} {\bibinfo {author} {\bibfnamefont {J.}~\bibnamefont
  {Toner}}, \bibinfo {author} {\bibfnamefont {Y.}~\bibnamefont {Tu}}, \ and\
  \bibinfo {author} {\bibfnamefont {S.}~\bibnamefont {Ramaswamy}},\ }\href@noop
  {} {\bibfield  {journal} {\bibinfo  {journal} {Annals of Physics}\ }\textbf
  {\bibinfo {volume} {318}},\ \bibinfo {pages} {170} (\bibinfo {year}
  {2005})}\BibitemShut {NoStop}%
\bibitem [{\citenamefont {Romanczuk}\ \emph {et~al.}(2012)\citenamefont
  {Romanczuk}, \citenamefont {Bär}, \citenamefont {Ebeling}, \citenamefont
  {Lindner},\ and\ \citenamefont {Schimansky-Geier}}]{Romanczuk_2012}%
  \BibitemOpen
  \bibfield  {author} {\bibinfo {author} {\bibfnamefont {P.}~\bibnamefont
  {Romanczuk}}, \bibinfo {author} {\bibfnamefont {M.}~\bibnamefont {Bär}},
  \bibinfo {author} {\bibfnamefont {W.}~\bibnamefont {Ebeling}}, \bibinfo
  {author} {\bibfnamefont {B.}~\bibnamefont {Lindner}}, \ and\ \bibinfo
  {author} {\bibfnamefont {L.}~\bibnamefont {Schimansky-Geier}},\ }\href
  {\doibase 10.1140/epjst/e2012-01529-y} {\bibfield  {journal} {\bibinfo
  {journal} {The European Physical Journal Special Topics}\ }\textbf {\bibinfo
  {volume} {202}},\ \bibinfo {pages} {1–162} (\bibinfo {year}
  {2012})}\BibitemShut {NoStop}%
\bibitem [{\citenamefont {Prost}\ \emph {et~al.}(2015)\citenamefont {Prost},
  \citenamefont {J{\"u}licher},\ and\ \citenamefont
  {Joanny}}]{prost2015active}%
  \BibitemOpen
  \bibfield  {author} {\bibinfo {author} {\bibfnamefont {J.}~\bibnamefont
  {Prost}}, \bibinfo {author} {\bibfnamefont {F.}~\bibnamefont {J{\"u}licher}},
  \ and\ \bibinfo {author} {\bibfnamefont {J.-F.}\ \bibnamefont {Joanny}},\
  }\href@noop {} {\bibfield  {journal} {\bibinfo  {journal} {Nature physics}\
  }\textbf {\bibinfo {volume} {11}},\ \bibinfo {pages} {111} (\bibinfo {year}
  {2015})}\BibitemShut {NoStop}%
\bibitem [{\citenamefont {Ramaswamy}(2019)}]{ramaswamy2019active}%
  \BibitemOpen
  \bibfield  {author} {\bibinfo {author} {\bibfnamefont {S.}~\bibnamefont
  {Ramaswamy}},\ }\href@noop {} {\bibfield  {journal} {\bibinfo  {journal}
  {Nature Reviews Physics}\ }\textbf {\bibinfo {volume} {1}},\ \bibinfo {pages}
  {640} (\bibinfo {year} {2019})}\BibitemShut {NoStop}%
\bibitem [{\citenamefont {Gompper}\ \emph {et~al.}(2020)\citenamefont
  {Gompper}, \citenamefont {Winkler}, \citenamefont {Speck}, \citenamefont
  {Solon}, \citenamefont {Nardini}, \citenamefont {Peruani}, \citenamefont
  {L{\"o}wen}, \citenamefont {Golestanian}, \citenamefont {Kaupp},
  \citenamefont {Alvarez} \emph {et~al.}}]{gompper20202020}%
  \BibitemOpen
  \bibfield  {author} {\bibinfo {author} {\bibfnamefont {G.}~\bibnamefont
  {Gompper}}, \bibinfo {author} {\bibfnamefont {R.~G.}\ \bibnamefont
  {Winkler}}, \bibinfo {author} {\bibfnamefont {T.}~\bibnamefont {Speck}},
  \bibinfo {author} {\bibfnamefont {A.}~\bibnamefont {Solon}}, \bibinfo
  {author} {\bibfnamefont {C.}~\bibnamefont {Nardini}}, \bibinfo {author}
  {\bibfnamefont {F.}~\bibnamefont {Peruani}}, \bibinfo {author} {\bibfnamefont
  {H.}~\bibnamefont {L{\"o}wen}}, \bibinfo {author} {\bibfnamefont
  {R.}~\bibnamefont {Golestanian}}, \bibinfo {author} {\bibfnamefont {U.~B.}\
  \bibnamefont {Kaupp}}, \bibinfo {author} {\bibfnamefont {L.}~\bibnamefont
  {Alvarez}},  \emph {et~al.},\ }\href@noop {} {\bibfield  {journal} {\bibinfo
  {journal} {Journal of Physics: Condensed Matter}\ }\textbf {\bibinfo {volume}
  {32}},\ \bibinfo {pages} {193001} (\bibinfo {year} {2020})}\BibitemShut
  {NoStop}%
\bibitem [{\citenamefont {Aditi~Simha}\ and\ \citenamefont
  {Ramaswamy}(2002)}]{PhysRevLett.89.058101}%
  \BibitemOpen
  \bibfield  {author} {\bibinfo {author} {\bibfnamefont {R.}~\bibnamefont
  {Aditi~Simha}}\ and\ \bibinfo {author} {\bibfnamefont {S.}~\bibnamefont
  {Ramaswamy}},\ }\href {\doibase 10.1103/PhysRevLett.89.058101} {\bibfield
  {journal} {\bibinfo  {journal} {Phys. Rev. Lett.}\ }\textbf {\bibinfo
  {volume} {89}},\ \bibinfo {pages} {058101} (\bibinfo {year}
  {2002})}\BibitemShut {NoStop}%
\bibitem [{\citenamefont {Bowick}\ \emph {et~al.}(2022)\citenamefont {Bowick},
  \citenamefont {Fakhri}, \citenamefont {Marchetti},\ and\ \citenamefont
  {Ramaswamy}}]{PhysRevX.12.010501}%
  \BibitemOpen
  \bibfield  {author} {\bibinfo {author} {\bibfnamefont {M.~J.}\ \bibnamefont
  {Bowick}}, \bibinfo {author} {\bibfnamefont {N.}~\bibnamefont {Fakhri}},
  \bibinfo {author} {\bibfnamefont {M.~C.}\ \bibnamefont {Marchetti}}, \ and\
  \bibinfo {author} {\bibfnamefont {S.}~\bibnamefont {Ramaswamy}},\ }\href
  {\doibase 10.1103/PhysRevX.12.010501} {\bibfield  {journal} {\bibinfo
  {journal} {Phys. Rev. X}\ }\textbf {\bibinfo {volume} {12}},\ \bibinfo
  {pages} {010501} (\bibinfo {year} {2022})}\BibitemShut {NoStop}%
\bibitem [{\citenamefont {Wensink}\ \emph {et~al.}(2012)\citenamefont
  {Wensink}, \citenamefont {Dunkel}, \citenamefont {Heidenreich}, \citenamefont
  {Drescher}, \citenamefont {Goldstein}, \citenamefont {L{\"o}wen},\ and\
  \citenamefont {Yeomans}}]{Yeomans-PNAS-2012}%
  \BibitemOpen
  \bibfield  {author} {\bibinfo {author} {\bibfnamefont {H.~H.}\ \bibnamefont
  {Wensink}}, \bibinfo {author} {\bibfnamefont {J.}~\bibnamefont {Dunkel}},
  \bibinfo {author} {\bibfnamefont {S.}~\bibnamefont {Heidenreich}}, \bibinfo
  {author} {\bibfnamefont {K.}~\bibnamefont {Drescher}}, \bibinfo {author}
  {\bibfnamefont {R.~E.}\ \bibnamefont {Goldstein}}, \bibinfo {author}
  {\bibfnamefont {H.}~\bibnamefont {L{\"o}wen}}, \ and\ \bibinfo {author}
  {\bibfnamefont {J.~M.}\ \bibnamefont {Yeomans}},\ }\href@noop {} {\bibfield
  {journal} {\bibinfo  {journal} {Proceedings of the national academy of
  sciences}\ }\textbf {\bibinfo {volume} {109}},\ \bibinfo {pages} {14308}
  (\bibinfo {year} {2012})}\BibitemShut {NoStop}%
\bibitem [{\citenamefont {Doostmohammadi}\ \emph {et~al.}(2018)\citenamefont
  {Doostmohammadi}, \citenamefont {Ign{\'e}s-Mullol}, \citenamefont {Yeomans},\
  and\ \citenamefont {Sagu{\'e}s}}]{Yeomans-Natcomm-2018}%
  \BibitemOpen
  \bibfield  {author} {\bibinfo {author} {\bibfnamefont {A.}~\bibnamefont
  {Doostmohammadi}}, \bibinfo {author} {\bibfnamefont {J.}~\bibnamefont
  {Ign{\'e}s-Mullol}}, \bibinfo {author} {\bibfnamefont {J.~M.}\ \bibnamefont
  {Yeomans}}, \ and\ \bibinfo {author} {\bibfnamefont {F.}~\bibnamefont
  {Sagu{\'e}s}},\ }\href@noop {} {\bibfield  {journal} {\bibinfo  {journal}
  {Nature communications}\ }\textbf {\bibinfo {volume} {9}},\ \bibinfo {pages}
  {1} (\bibinfo {year} {2018})}\BibitemShut {NoStop}%
\bibitem [{\citenamefont {Thijssen}\ \emph {et~al.}(2020)\citenamefont
  {Thijssen}, \citenamefont {Nejad},\ and\ \citenamefont
  {Yeomans}}]{Yeomans-PRL-2020}%
  \BibitemOpen
  \bibfield  {author} {\bibinfo {author} {\bibfnamefont {K.}~\bibnamefont
  {Thijssen}}, \bibinfo {author} {\bibfnamefont {M.~R.}\ \bibnamefont {Nejad}},
  \ and\ \bibinfo {author} {\bibfnamefont {J.~M.}\ \bibnamefont {Yeomans}},\
  }\href {\doibase 10.1103/PhysRevLett.125.218004} {\bibfield  {journal}
  {\bibinfo  {journal} {Phys. Rev. Lett.}\ }\textbf {\bibinfo {volume} {125}},\
  \bibinfo {pages} {218004} (\bibinfo {year} {2020})}\BibitemShut {NoStop}%
\bibitem [{\citenamefont {Maitra}\ and\ \citenamefont
  {Voituriez}(2020)}]{AnanyoPRL2020}%
  \BibitemOpen
  \bibfield  {author} {\bibinfo {author} {\bibfnamefont {A.}~\bibnamefont
  {Maitra}}\ and\ \bibinfo {author} {\bibfnamefont {R.}~\bibnamefont
  {Voituriez}},\ }\href {\doibase 10.1103/PhysRevLett.124.048003} {\bibfield
  {journal} {\bibinfo  {journal} {Phys. Rev. Lett.}\ }\textbf {\bibinfo
  {volume} {124}},\ \bibinfo {pages} {048003} (\bibinfo {year}
  {2020})}\BibitemShut {NoStop}%
\bibitem [{\citenamefont {Nejad}\ and\ \citenamefont
  {Yeomans}(2022)}]{Yeomans-PRL-2022}%
  \BibitemOpen
  \bibfield  {author} {\bibinfo {author} {\bibfnamefont {M.~R.}\ \bibnamefont
  {Nejad}}\ and\ \bibinfo {author} {\bibfnamefont {J.~M.}\ \bibnamefont
  {Yeomans}},\ }\href {\doibase 10.1103/PhysRevLett.128.048001} {\bibfield
  {journal} {\bibinfo  {journal} {Phys. Rev. Lett.}\ }\textbf {\bibinfo
  {volume} {128}},\ \bibinfo {pages} {048001} (\bibinfo {year}
  {2022})}\BibitemShut {NoStop}%
\bibitem [{\citenamefont {Buttinoni}\ \emph {et~al.}(2013)\citenamefont
  {Buttinoni}, \citenamefont {Bialk\'e}, \citenamefont {K\"ummel},
  \citenamefont {L\"owen}, \citenamefont {Bechinger},\ and\ \citenamefont
  {Speck}}]{Hartmut2013}%
  \BibitemOpen
  \bibfield  {author} {\bibinfo {author} {\bibfnamefont {I.}~\bibnamefont
  {Buttinoni}}, \bibinfo {author} {\bibfnamefont {J.}~\bibnamefont {Bialk\'e}},
  \bibinfo {author} {\bibfnamefont {F.}~\bibnamefont {K\"ummel}}, \bibinfo
  {author} {\bibfnamefont {H.}~\bibnamefont {L\"owen}}, \bibinfo {author}
  {\bibfnamefont {C.}~\bibnamefont {Bechinger}}, \ and\ \bibinfo {author}
  {\bibfnamefont {T.}~\bibnamefont {Speck}},\ }\href {\doibase
  10.1103/PhysRevLett.110.238301} {\bibfield  {journal} {\bibinfo  {journal}
  {Phys. Rev. Lett.}\ }\textbf {\bibinfo {volume} {110}},\ \bibinfo {pages}
  {238301} (\bibinfo {year} {2013})}\BibitemShut {NoStop}%
\bibitem [{\citenamefont {Bialk\'e}\ \emph {et~al.}(2015)\citenamefont
  {Bialk\'e}, \citenamefont {Siebert}, \citenamefont {L\"owen},\ and\
  \citenamefont {Speck}}]{Hartmut2015}%
  \BibitemOpen
  \bibfield  {author} {\bibinfo {author} {\bibfnamefont {J.}~\bibnamefont
  {Bialk\'e}}, \bibinfo {author} {\bibfnamefont {J.~T.}\ \bibnamefont
  {Siebert}}, \bibinfo {author} {\bibfnamefont {H.}~\bibnamefont {L\"owen}}, \
  and\ \bibinfo {author} {\bibfnamefont {T.}~\bibnamefont {Speck}},\ }\href
  {\doibase 10.1103/PhysRevLett.115.098301} {\bibfield  {journal} {\bibinfo
  {journal} {Phys. Rev. Lett.}\ }\textbf {\bibinfo {volume} {115}},\ \bibinfo
  {pages} {098301} (\bibinfo {year} {2015})}\BibitemShut {NoStop}%
\bibitem [{\citenamefont {Ivlev}\ \emph {et~al.}(2015)\citenamefont {Ivlev},
  \citenamefont {Bartnick}, \citenamefont {Heinen}, \citenamefont {Du},
  \citenamefont {Nosenko},\ and\ \citenamefont {L\"owen}}]{Hartmut-PRX-2015}%
  \BibitemOpen
  \bibfield  {author} {\bibinfo {author} {\bibfnamefont {A.~V.}\ \bibnamefont
  {Ivlev}}, \bibinfo {author} {\bibfnamefont {J.}~\bibnamefont {Bartnick}},
  \bibinfo {author} {\bibfnamefont {M.}~\bibnamefont {Heinen}}, \bibinfo
  {author} {\bibfnamefont {C.-R.}\ \bibnamefont {Du}}, \bibinfo {author}
  {\bibfnamefont {V.}~\bibnamefont {Nosenko}}, \ and\ \bibinfo {author}
  {\bibfnamefont {H.}~\bibnamefont {L\"owen}},\ }\href {\doibase
  10.1103/PhysRevX.5.011035} {\bibfield  {journal} {\bibinfo  {journal} {Phys.
  Rev. X}\ }\textbf {\bibinfo {volume} {5}},\ \bibinfo {pages} {011035}
  (\bibinfo {year} {2015})}\BibitemShut {NoStop}%
\bibitem [{\citenamefont {Bechinger}\ \emph {et~al.}(2016)\citenamefont
  {Bechinger}, \citenamefont {Di~Leonardo}, \citenamefont {L\"owen},
  \citenamefont {Reichhardt}, \citenamefont {Volpe},\ and\ \citenamefont
  {Volpe}}]{Hartmut2016}%
  \BibitemOpen
  \bibfield  {author} {\bibinfo {author} {\bibfnamefont {C.}~\bibnamefont
  {Bechinger}}, \bibinfo {author} {\bibfnamefont {R.}~\bibnamefont
  {Di~Leonardo}}, \bibinfo {author} {\bibfnamefont {H.}~\bibnamefont
  {L\"owen}}, \bibinfo {author} {\bibfnamefont {C.}~\bibnamefont {Reichhardt}},
  \bibinfo {author} {\bibfnamefont {G.}~\bibnamefont {Volpe}}, \ and\ \bibinfo
  {author} {\bibfnamefont {G.}~\bibnamefont {Volpe}},\ }\href {\doibase
  10.1103/RevModPhys.88.045006} {\bibfield  {journal} {\bibinfo  {journal}
  {Rev. Mod. Phys.}\ }\textbf {\bibinfo {volume} {88}},\ \bibinfo {pages}
  {045006} (\bibinfo {year} {2016})}\BibitemShut {NoStop}%
\bibitem [{\citenamefont {Mandal}\ \emph {et~al.}(2019)\citenamefont {Mandal},
  \citenamefont {Liebchen},\ and\ \citenamefont {L\"owen}}]{Hartmut2019}%
  \BibitemOpen
  \bibfield  {author} {\bibinfo {author} {\bibfnamefont {S.}~\bibnamefont
  {Mandal}}, \bibinfo {author} {\bibfnamefont {B.}~\bibnamefont {Liebchen}}, \
  and\ \bibinfo {author} {\bibfnamefont {H.}~\bibnamefont {L\"owen}},\ }\href
  {\doibase 10.1103/PhysRevLett.123.228001} {\bibfield  {journal} {\bibinfo
  {journal} {Phys. Rev. Lett.}\ }\textbf {\bibinfo {volume} {123}},\ \bibinfo
  {pages} {228001} (\bibinfo {year} {2019})}\BibitemShut {NoStop}%
\bibitem [{\citenamefont {Löwen}(2020)}]{Hartmut2020}%
  \BibitemOpen
  \bibfield  {author} {\bibinfo {author} {\bibfnamefont {H.}~\bibnamefont
  {Löwen}},\ }\href {\doibase 10.1063/1.5134455} {\bibfield  {journal}
  {\bibinfo  {journal} {The Journal of Chemical Physics}\ }\textbf {\bibinfo
  {volume} {152}},\ \bibinfo {pages} {040901} (\bibinfo {year}
  {2020})}\BibitemShut {NoStop}%
\bibitem [{\citenamefont {Speck}\ \emph {et~al.}(2014)\citenamefont {Speck},
  \citenamefont {Bialk\'e}, \citenamefont {Menzel},\ and\ \citenamefont
  {L\"owen}}]{Speck2014}%
  \BibitemOpen
  \bibfield  {author} {\bibinfo {author} {\bibfnamefont {T.}~\bibnamefont
  {Speck}}, \bibinfo {author} {\bibfnamefont {J.}~\bibnamefont {Bialk\'e}},
  \bibinfo {author} {\bibfnamefont {A.~M.}\ \bibnamefont {Menzel}}, \ and\
  \bibinfo {author} {\bibfnamefont {H.}~\bibnamefont {L\"owen}},\ }\href
  {\doibase 10.1103/PhysRevLett.112.218304} {\bibfield  {journal} {\bibinfo
  {journal} {Phys. Rev. Lett.}\ }\textbf {\bibinfo {volume} {112}},\ \bibinfo
  {pages} {218304} (\bibinfo {year} {2014})}\BibitemShut {NoStop}%
\bibitem [{\citenamefont {McCandlish}\ \emph {et~al.}(2012)\citenamefont
  {McCandlish}, \citenamefont {Baskaran},\ and\ \citenamefont
  {Hagan}}]{Baskaran-softmatter}%
  \BibitemOpen
  \bibfield  {author} {\bibinfo {author} {\bibfnamefont {S.~R.}\ \bibnamefont
  {McCandlish}}, \bibinfo {author} {\bibfnamefont {A.}~\bibnamefont
  {Baskaran}}, \ and\ \bibinfo {author} {\bibfnamefont {M.~F.}\ \bibnamefont
  {Hagan}},\ }\href@noop {} {\bibfield  {journal} {\bibinfo  {journal} {Soft
  Matter}\ }\textbf {\bibinfo {volume} {8}},\ \bibinfo {pages} {2527} (\bibinfo
  {year} {2012})}\BibitemShut {NoStop}%
\bibitem [{\citenamefont {Redner}\ \emph {et~al.}(2013)\citenamefont {Redner},
  \citenamefont {Hagan},\ and\ \citenamefont {Baskaran}}]{redner-baskaran}%
  \BibitemOpen
  \bibfield  {author} {\bibinfo {author} {\bibfnamefont {G.~S.}\ \bibnamefont
  {Redner}}, \bibinfo {author} {\bibfnamefont {M.~F.}\ \bibnamefont {Hagan}}, \
  and\ \bibinfo {author} {\bibfnamefont {A.}~\bibnamefont {Baskaran}},\ }\href
  {\doibase 10.1103/PhysRevLett.110.055701} {\bibfield  {journal} {\bibinfo
  {journal} {Phys. Rev. Lett.}\ }\textbf {\bibinfo {volume} {110}},\ \bibinfo
  {pages} {055701} (\bibinfo {year} {2013})}\BibitemShut {NoStop}%
\bibitem [{\citenamefont {Vicsek}\ and\ \citenamefont
  {Zafeiris}(2012)}]{vicsek2012}%
  \BibitemOpen
  \bibfield  {author} {\bibinfo {author} {\bibfnamefont {T.}~\bibnamefont
  {Vicsek}}\ and\ \bibinfo {author} {\bibfnamefont {A.}~\bibnamefont
  {Zafeiris}},\ }\href@noop {} {\bibfield  {journal} {\bibinfo  {journal}
  {Physics reports}\ }\textbf {\bibinfo {volume} {517}},\ \bibinfo {pages} {71}
  (\bibinfo {year} {2012})}\BibitemShut {NoStop}%
\bibitem [{\citenamefont {Toner}\ and\ \citenamefont
  {Tu}(1995)}]{toner-Tu-1995}%
  \BibitemOpen
  \bibfield  {author} {\bibinfo {author} {\bibfnamefont {J.}~\bibnamefont
  {Toner}}\ and\ \bibinfo {author} {\bibfnamefont {Y.}~\bibnamefont {Tu}},\
  }\href {\doibase 10.1103/PhysRevLett.75.4326} {\bibfield  {journal} {\bibinfo
   {journal} {Phys. Rev. Lett.}\ }\textbf {\bibinfo {volume} {75}},\ \bibinfo
  {pages} {4326} (\bibinfo {year} {1995})}\BibitemShut {NoStop}%
\bibitem [{\citenamefont {Wittkowski}\ \emph {et~al.}(2014)\citenamefont
  {Wittkowski}, \citenamefont {Tiribocchi}, \citenamefont {Stenhammar},
  \citenamefont {Allen}, \citenamefont {Marenduzzo},\ and\ \citenamefont
  {Cates}}]{wittkowski2014scalar}%
  \BibitemOpen
  \bibfield  {author} {\bibinfo {author} {\bibfnamefont {R.}~\bibnamefont
  {Wittkowski}}, \bibinfo {author} {\bibfnamefont {A.}~\bibnamefont
  {Tiribocchi}}, \bibinfo {author} {\bibfnamefont {J.}~\bibnamefont
  {Stenhammar}}, \bibinfo {author} {\bibfnamefont {R.~J.}\ \bibnamefont
  {Allen}}, \bibinfo {author} {\bibfnamefont {D.}~\bibnamefont {Marenduzzo}}, \
  and\ \bibinfo {author} {\bibfnamefont {M.~E.}\ \bibnamefont {Cates}},\
  }\href@noop {} {\bibfield  {journal} {\bibinfo  {journal} {Nature
  communications}\ }\textbf {\bibinfo {volume} {5}},\ \bibinfo {pages} {1}
  (\bibinfo {year} {2014})}\BibitemShut {NoStop}%
\bibitem [{\citenamefont {Saha}\ \emph {et~al.}(2020)\citenamefont {Saha},
  \citenamefont {Agudo-Canalejo},\ and\ \citenamefont
  {Golestanian}}]{Golestanian-PRX-2020}%
  \BibitemOpen
  \bibfield  {author} {\bibinfo {author} {\bibfnamefont {S.}~\bibnamefont
  {Saha}}, \bibinfo {author} {\bibfnamefont {J.}~\bibnamefont
  {Agudo-Canalejo}}, \ and\ \bibinfo {author} {\bibfnamefont {R.}~\bibnamefont
  {Golestanian}},\ }\href {\doibase 10.1103/PhysRevX.10.041009} {\bibfield
  {journal} {\bibinfo  {journal} {Phys. Rev. X}\ }\textbf {\bibinfo {volume}
  {10}},\ \bibinfo {pages} {041009} (\bibinfo {year} {2020})}\BibitemShut
  {NoStop}%
\bibitem [{\citenamefont {Grosberg}\ and\ \citenamefont
  {Joanny}(2015)}]{joanny-2015}%
  \BibitemOpen
  \bibfield  {author} {\bibinfo {author} {\bibfnamefont {A.~Y.}\ \bibnamefont
  {Grosberg}}\ and\ \bibinfo {author} {\bibfnamefont {J.-F.}\ \bibnamefont
  {Joanny}},\ }\href {\doibase 10.1103/PhysRevE.92.032118} {\bibfield
  {journal} {\bibinfo  {journal} {Phys. Rev. E}\ }\textbf {\bibinfo {volume}
  {92}},\ \bibinfo {pages} {032118} (\bibinfo {year} {2015})}\BibitemShut
  {NoStop}%
\bibitem [{\citenamefont {Weber}\ \emph {et~al.}(2016)\citenamefont {Weber},
  \citenamefont {Weber},\ and\ \citenamefont {Frey}}]{frey}%
  \BibitemOpen
  \bibfield  {author} {\bibinfo {author} {\bibfnamefont {S.~N.}\ \bibnamefont
  {Weber}}, \bibinfo {author} {\bibfnamefont {C.~A.}\ \bibnamefont {Weber}}, \
  and\ \bibinfo {author} {\bibfnamefont {E.}~\bibnamefont {Frey}},\ }\href
  {\doibase 10.1103/PhysRevLett.116.058301} {\bibfield  {journal} {\bibinfo
  {journal} {Phys. Rev. Lett.}\ }\textbf {\bibinfo {volume} {116}},\ \bibinfo
  {pages} {058301} (\bibinfo {year} {2016})}\BibitemShut {NoStop}%
\bibitem [{\citenamefont {Hohenberg}\ and\ \citenamefont
  {Halperin}(1977)}]{RevModPhys.49.435}%
  \BibitemOpen
  \bibfield  {author} {\bibinfo {author} {\bibfnamefont {P.~C.}\ \bibnamefont
  {Hohenberg}}\ and\ \bibinfo {author} {\bibfnamefont {B.~I.}\ \bibnamefont
  {Halperin}},\ }\href {\doibase 10.1103/RevModPhys.49.435} {\bibfield
  {journal} {\bibinfo  {journal} {Rev. Mod. Phys.}\ }\textbf {\bibinfo {volume}
  {49}},\ \bibinfo {pages} {435} (\bibinfo {year} {1977})}\BibitemShut
  {NoStop}%
\bibitem [{\citenamefont {Grosberg}\ and\ \citenamefont
  {Joanny}(2018)}]{joanny-2018}%
  \BibitemOpen
  \bibfield  {author} {\bibinfo {author} {\bibfnamefont {A.~Y.}\ \bibnamefont
  {Grosberg}}\ and\ \bibinfo {author} {\bibfnamefont {J.-F.}\ \bibnamefont
  {Joanny}},\ }\href@noop {} {\bibfield  {journal} {\bibinfo  {journal}
  {Polymer Science, Series C}\ }\textbf {\bibinfo {volume} {60}},\ \bibinfo
  {pages} {118} (\bibinfo {year} {2018})}\BibitemShut {NoStop}%
\bibitem [{\citenamefont {Ilker}\ and\ \citenamefont
  {Joanny}(2020)}]{joanny-2020}%
  \BibitemOpen
  \bibfield  {author} {\bibinfo {author} {\bibfnamefont {E.}~\bibnamefont
  {Ilker}}\ and\ \bibinfo {author} {\bibfnamefont {J.-F.}\ \bibnamefont
  {Joanny}},\ }\href {\doibase 10.1103/PhysRevResearch.2.023200} {\bibfield
  {journal} {\bibinfo  {journal} {Phys. Rev. Research}\ }\textbf {\bibinfo
  {volume} {2}},\ \bibinfo {pages} {023200} (\bibinfo {year}
  {2020})}\BibitemShut {NoStop}%
\bibitem [{\citenamefont {Ganai}\ \emph {et~al.}(2014)\citenamefont {Ganai},
  \citenamefont {Sengupta},\ and\ \citenamefont {Menon}}]{ganai}%
  \BibitemOpen
  \bibfield  {author} {\bibinfo {author} {\bibfnamefont {N.}~\bibnamefont
  {Ganai}}, \bibinfo {author} {\bibfnamefont {S.}~\bibnamefont {Sengupta}}, \
  and\ \bibinfo {author} {\bibfnamefont {G.~I.}\ \bibnamefont {Menon}},\
  }\href@noop {} {\bibfield  {journal} {\bibinfo  {journal} {Nucleic acids
  research}\ }\textbf {\bibinfo {volume} {42}},\ \bibinfo {pages} {4145}
  (\bibinfo {year} {2014})}\BibitemShut {NoStop}%
\bibitem [{\citenamefont {Smrek}\ and\ \citenamefont {Kremer}(2017)}]{kremer1}%
  \BibitemOpen
  \bibfield  {author} {\bibinfo {author} {\bibfnamefont {J.}~\bibnamefont
  {Smrek}}\ and\ \bibinfo {author} {\bibfnamefont {K.}~\bibnamefont {Kremer}},\
  }\href {\doibase 10.1103/PhysRevLett.118.098002} {\bibfield  {journal}
  {\bibinfo  {journal} {Phys. Rev. Lett.}\ }\textbf {\bibinfo {volume} {118}},\
  \bibinfo {pages} {098002} (\bibinfo {year} {2017})}\BibitemShut {NoStop}%
\bibitem [{\citenamefont {Smrek}\ and\ \citenamefont {Kremer}(2018)}]{kremer2}%
  \BibitemOpen
  \bibfield  {author} {\bibinfo {author} {\bibfnamefont {J.}~\bibnamefont
  {Smrek}}\ and\ \bibinfo {author} {\bibfnamefont {K.}~\bibnamefont {Kremer}},\
  }\href@noop {} {\bibfield  {journal} {\bibinfo  {journal} {Entropy}\ }\textbf
  {\bibinfo {volume} {20}},\ \bibinfo {pages} {520} (\bibinfo {year}
  {2018})}\BibitemShut {NoStop}%
\bibitem [{\citenamefont {Chari}\ \emph {et~al.}(2019)\citenamefont {Chari},
  \citenamefont {Dasgupta},\ and\ \citenamefont {Maiti}}]{Siva}%
  \BibitemOpen
  \bibfield  {author} {\bibinfo {author} {\bibfnamefont {S.~S.~N.}\
  \bibnamefont {Chari}}, \bibinfo {author} {\bibfnamefont {C.}~\bibnamefont
  {Dasgupta}}, \ and\ \bibinfo {author} {\bibfnamefont {P.~K.}\ \bibnamefont
  {Maiti}},\ }\href@noop {} {\bibfield  {journal} {\bibinfo  {journal} {Soft
  matter}\ }\textbf {\bibinfo {volume} {15}},\ \bibinfo {pages} {7275}
  (\bibinfo {year} {2019})}\BibitemShut {NoStop}%
\bibitem [{\citenamefont {Chattopadhyay}\ \emph {et~al.}(2021)\citenamefont
  {Chattopadhyay}, \citenamefont {Pannir-Sivajothi}, \citenamefont {Varma},
  \citenamefont {Ramaswamy}, \citenamefont {Dasgupta},\ and\ \citenamefont
  {Maiti}}]{Active-SRS}%
  \BibitemOpen
  \bibfield  {author} {\bibinfo {author} {\bibfnamefont {J.}~\bibnamefont
  {Chattopadhyay}}, \bibinfo {author} {\bibfnamefont {S.}~\bibnamefont
  {Pannir-Sivajothi}}, \bibinfo {author} {\bibfnamefont {K.}~\bibnamefont
  {Varma}}, \bibinfo {author} {\bibfnamefont {S.}~\bibnamefont {Ramaswamy}},
  \bibinfo {author} {\bibfnamefont {C.}~\bibnamefont {Dasgupta}}, \ and\
  \bibinfo {author} {\bibfnamefont {P.~K.}\ \bibnamefont {Maiti}},\ }\href
  {\doibase 10.1103/PhysRevE.104.054610} {\bibfield  {journal} {\bibinfo
  {journal} {Phys. Rev. E}\ }\textbf {\bibinfo {volume} {104}},\ \bibinfo
  {pages} {054610} (\bibinfo {year} {2021})}\BibitemShut {NoStop}%
\bibitem [{\citenamefont {V}\ \emph {et~al.}(2021)\citenamefont {V},
  \citenamefont {Lin},\ and\ \citenamefont
  {Maiti}}]{https://doi.org/10.48550/arxiv.2109.00415}%
  \BibitemOpen
  \bibfield  {author} {\bibinfo {author} {\bibfnamefont {N.}~\bibnamefont {V}},
  \bibinfo {author} {\bibfnamefont {S.-T.}\ \bibnamefont {Lin}}, \ and\
  \bibinfo {author} {\bibfnamefont {P.~K.}\ \bibnamefont {Maiti}},\ }\href
  {\doibase 10.48550/ARXIV.2109.00415} {\enquote {\bibinfo {title} {Phase
  behavior of active and passive dumbbells},}\ } (\bibinfo {year}
  {2021})\BibitemShut {NoStop}%
\bibitem [{\citenamefont {Allen}\ \emph {et~al.}(1993)\citenamefont {Allen},
  \citenamefont {Evans}, \citenamefont {Frenkel},\ and\ \citenamefont
  {Mulder}}]{allen1993hard}%
  \BibitemOpen
  \bibfield  {author} {\bibinfo {author} {\bibfnamefont {M.~P.}\ \bibnamefont
  {Allen}}, \bibinfo {author} {\bibfnamefont {G.~T.}\ \bibnamefont {Evans}},
  \bibinfo {author} {\bibfnamefont {D.}~\bibnamefont {Frenkel}}, \ and\
  \bibinfo {author} {\bibfnamefont {B.}~\bibnamefont {Mulder}},\ }\href@noop {}
  {\bibfield  {journal} {\bibinfo  {journal} {Advances in chemical physics}\
  }\textbf {\bibinfo {volume} {86}},\ \bibinfo {pages} {1} (\bibinfo {year}
  {1993})}\BibitemShut {NoStop}%
\bibitem [{\citenamefont {Vega}\ and\ \citenamefont
  {Lago}(1994)}]{vega1994fast}%
  \BibitemOpen
  \bibfield  {author} {\bibinfo {author} {\bibfnamefont {C.}~\bibnamefont
  {Vega}}\ and\ \bibinfo {author} {\bibfnamefont {S.}~\bibnamefont {Lago}},\
  }\href@noop {} {\bibfield  {journal} {\bibinfo  {journal} {Computers \&
  chemistry}\ }\textbf {\bibinfo {volume} {18}},\ \bibinfo {pages} {55}
  (\bibinfo {year} {1994})}\BibitemShut {NoStop}%
\bibitem [{\citenamefont {Earl}\ \emph {et~al.}(2001)\citenamefont {Earl},
  \citenamefont {Ilnytskyi},\ and\ \citenamefont {Wilson}}]{earl2001}%
  \BibitemOpen
  \bibfield  {author} {\bibinfo {author} {\bibfnamefont {D.~J.}\ \bibnamefont
  {Earl}}, \bibinfo {author} {\bibfnamefont {J.}~\bibnamefont {Ilnytskyi}}, \
  and\ \bibinfo {author} {\bibfnamefont {M.~R.}\ \bibnamefont {Wilson}},\
  }\href@noop {} {\bibfield  {journal} {\bibinfo  {journal} {Molecular
  physics}\ }\textbf {\bibinfo {volume} {99}},\ \bibinfo {pages} {1719}
  (\bibinfo {year} {2001})}\BibitemShut {NoStop}%
\bibitem [{\citenamefont {Weeks}\ \emph {et~al.}(1971)\citenamefont {Weeks},
  \citenamefont {Chandler},\ and\ \citenamefont {Andersen}}]{weeks1971}%
  \BibitemOpen
  \bibfield  {author} {\bibinfo {author} {\bibfnamefont {J.~D.}\ \bibnamefont
  {Weeks}}, \bibinfo {author} {\bibfnamefont {D.}~\bibnamefont {Chandler}}, \
  and\ \bibinfo {author} {\bibfnamefont {H.~C.}\ \bibnamefont {Andersen}},\
  }\href@noop {} {\bibfield  {journal} {\bibinfo  {journal} {The Journal of
  chemical physics}\ }\textbf {\bibinfo {volume} {54}},\ \bibinfo {pages}
  {5237} (\bibinfo {year} {1971})}\BibitemShut {NoStop}%
\bibitem [{\citenamefont {Rotunno}\ \emph {et~al.}(2004)\citenamefont
  {Rotunno}, \citenamefont {Bellini}, \citenamefont {Lansac},\ and\
  \citenamefont {Glaser}}]{rotunno2004}%
  \BibitemOpen
  \bibfield  {author} {\bibinfo {author} {\bibfnamefont {M.}~\bibnamefont
  {Rotunno}}, \bibinfo {author} {\bibfnamefont {T.}~\bibnamefont {Bellini}},
  \bibinfo {author} {\bibfnamefont {Y.}~\bibnamefont {Lansac}}, \ and\ \bibinfo
  {author} {\bibfnamefont {M.~A.}\ \bibnamefont {Glaser}},\ }\href@noop {}
  {\bibfield  {journal} {\bibinfo  {journal} {The Journal of chemical physics}\
  }\textbf {\bibinfo {volume} {121}},\ \bibinfo {pages} {5541} (\bibinfo {year}
  {2004})}\BibitemShut {NoStop}%
\bibitem [{\citenamefont {Berendsen}\ \emph {et~al.}(1984)\citenamefont
  {Berendsen}, \citenamefont {Postma}, \citenamefont {van Gunsteren},
  \citenamefont {DiNola},\ and\ \citenamefont {Haak}}]{berendsen1984}%
  \BibitemOpen
  \bibfield  {author} {\bibinfo {author} {\bibfnamefont {H.~J.}\ \bibnamefont
  {Berendsen}}, \bibinfo {author} {\bibfnamefont {J.~v.}\ \bibnamefont
  {Postma}}, \bibinfo {author} {\bibfnamefont {W.~F.}\ \bibnamefont {van
  Gunsteren}}, \bibinfo {author} {\bibfnamefont {A.}~\bibnamefont {DiNola}}, \
  and\ \bibinfo {author} {\bibfnamefont {J.~R.}\ \bibnamefont {Haak}},\
  }\href@noop {} {\bibfield  {journal} {\bibinfo  {journal} {The Journal of
  chemical physics}\ }\textbf {\bibinfo {volume} {81}},\ \bibinfo {pages}
  {3684} (\bibinfo {year} {1984})}\BibitemShut {NoStop}%
\bibitem [{sp()}]{sp}%
  \BibitemOpen
  \href@noop {} {\enquote {\bibinfo {title} {Supplemental information},}\
  }\BibitemShut {NoStop}%
\bibitem [{\citenamefont {Cuetos}\ \emph {et~al.}(2005)\citenamefont {Cuetos},
  \citenamefont {Martinez-Haya}, \citenamefont {Lago},\ and\ \citenamefont
  {Rull}}]{cuetos2005parsons}%
  \BibitemOpen
  \bibfield  {author} {\bibinfo {author} {\bibfnamefont {A.}~\bibnamefont
  {Cuetos}}, \bibinfo {author} {\bibfnamefont {B.}~\bibnamefont
  {Martinez-Haya}}, \bibinfo {author} {\bibfnamefont {S.}~\bibnamefont {Lago}},
  \ and\ \bibinfo {author} {\bibfnamefont {L.}~\bibnamefont {Rull}},\
  }\href@noop {} {\bibfield  {journal} {\bibinfo  {journal} {The Journal of
  Physical Chemistry B}\ }\textbf {\bibinfo {volume} {109}},\ \bibinfo {pages}
  {13729} (\bibinfo {year} {2005})}\BibitemShut {NoStop}%
\bibitem [{\citenamefont {Boublík}(1976)}]{boublik1976}%
  \BibitemOpen
  \bibfield  {author} {\bibinfo {author} {\bibfnamefont {T.}~\bibnamefont
  {Boublík}},\ }\href {\doibase 10.1080/00268977600103051} {\bibfield
  {journal} {\bibinfo  {journal} {Molecular Physics}\ }\textbf {\bibinfo
  {volume} {32}},\ \bibinfo {pages} {1737} (\bibinfo {year} {1976})},\ \Eprint
  {http://arxiv.org/abs/https://doi.org/10.1080/00268977600103051}
  {https://doi.org/10.1080/00268977600103051} \BibitemShut {NoStop}%
\bibitem [{\citenamefont {Hansen}\ and\ \citenamefont
  {McDonald}(2013)}]{hansen2013theory}%
  \BibitemOpen
  \bibfield  {author} {\bibinfo {author} {\bibfnamefont {J.-P.}\ \bibnamefont
  {Hansen}}\ and\ \bibinfo {author} {\bibfnamefont {I.~R.}\ \bibnamefont
  {McDonald}},\ }\href@noop {} {\emph {\bibinfo {title} {Theory of simple
  liquids: with applications to soft matter}}}\ (\bibinfo  {publisher}
  {Academic press},\ \bibinfo {year} {2013})\BibitemShut {NoStop}%
\bibitem [{\citenamefont {Bhuyan}\ \emph {et~al.}(2020)\citenamefont {Bhuyan},
  \citenamefont {Mandal}, \citenamefont {Chaudhuri}, \citenamefont {Dhar},\
  and\ \citenamefont {Dasgupta}}]{Pranab-PRE}%
  \BibitemOpen
  \bibfield  {author} {\bibinfo {author} {\bibfnamefont {P.~J.}\ \bibnamefont
  {Bhuyan}}, \bibinfo {author} {\bibfnamefont {R.}~\bibnamefont {Mandal}},
  \bibinfo {author} {\bibfnamefont {P.}~\bibnamefont {Chaudhuri}}, \bibinfo
  {author} {\bibfnamefont {A.}~\bibnamefont {Dhar}}, \ and\ \bibinfo {author}
  {\bibfnamefont {C.}~\bibnamefont {Dasgupta}},\ }\href {\doibase
  10.1103/PhysRevE.101.022125} {\bibfield  {journal} {\bibinfo  {journal}
  {Phys. Rev. E}\ }\textbf {\bibinfo {volume} {101}},\ \bibinfo {pages}
  {022125} (\bibinfo {year} {2020})}\BibitemShut {NoStop}%
\end{thebibliography}%

\appendix

\section{Appendixes}


\renewcommand{\thefigure}{S\arabic{figure}}
\setcounter{figure}{0}

\section{Pair correlation functions:}
To identify the ordered phases precisely, we calculate local nematic order parameter and relevant pair correlation functions in the phase separated regions.
We follow the usual way as defined by McGrother et.al \cite{McGrother}. Along with the radial distribution function $ g(r) $, we calculate orientational pair correlation function $ g_{2}(r) $ which is relevant for quantifying order-disorder transition. $ g_{2}(r) $ is defined as $ 2^{nd} $ order Legendre polynomial with the argument cosine of relative angle between spherocylinders $i$ and $j$. 
$ g_{2}(r) = P_{2}(\vec{u_{i}}.\vec{u_{j}})$. We further calculate 
vectorial pair correlation functions $ g_{\parallel}(r), g_{\perp}(r) $ which are the projection of radial distribution function $ g(r) $ along the directions parallel and perpendicular to the nematic director respectively. Periodic oscillations in $ g_{\parallel}(r) $ indicates presence of layering and thus distinguish between nematic and smectic phase.
In-layer periodicity is quantified by $ g_{\perp}(r) $ that differentiates between smectic and Crystal phase.

\begin{figure} [!htb]
	\centering
	\includegraphics[scale=0.5]{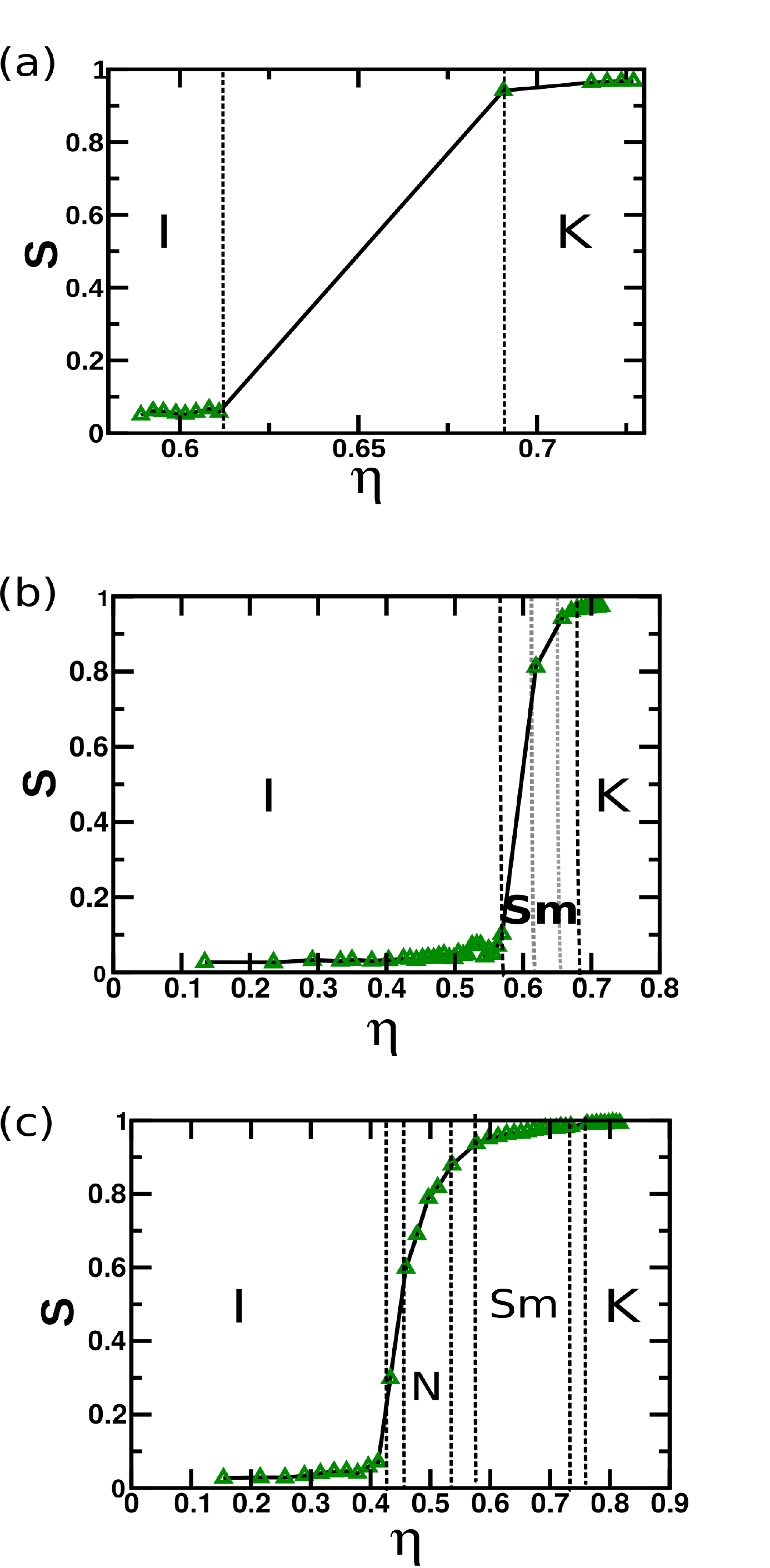}
	\caption{Equilibrium phase diagram of soft repulsive spherocylinders for the aspect ratios (a) $ L/D = 2$, (b) $3$ and (c) $5$ at $T^{*} = 5$. We plot packing fraction $\eta$ along x axis and nematic order parameter $ S $ along y axis. The vertical dashed lines show coexistence regions near the phase boundaries.
	} \label{eps-diffL}
	
\end{figure}

\section{Critical activity parameter for phase separation:}
\begin{figure*} [!htb]
	\centering
	\includegraphics[scale=0.9]{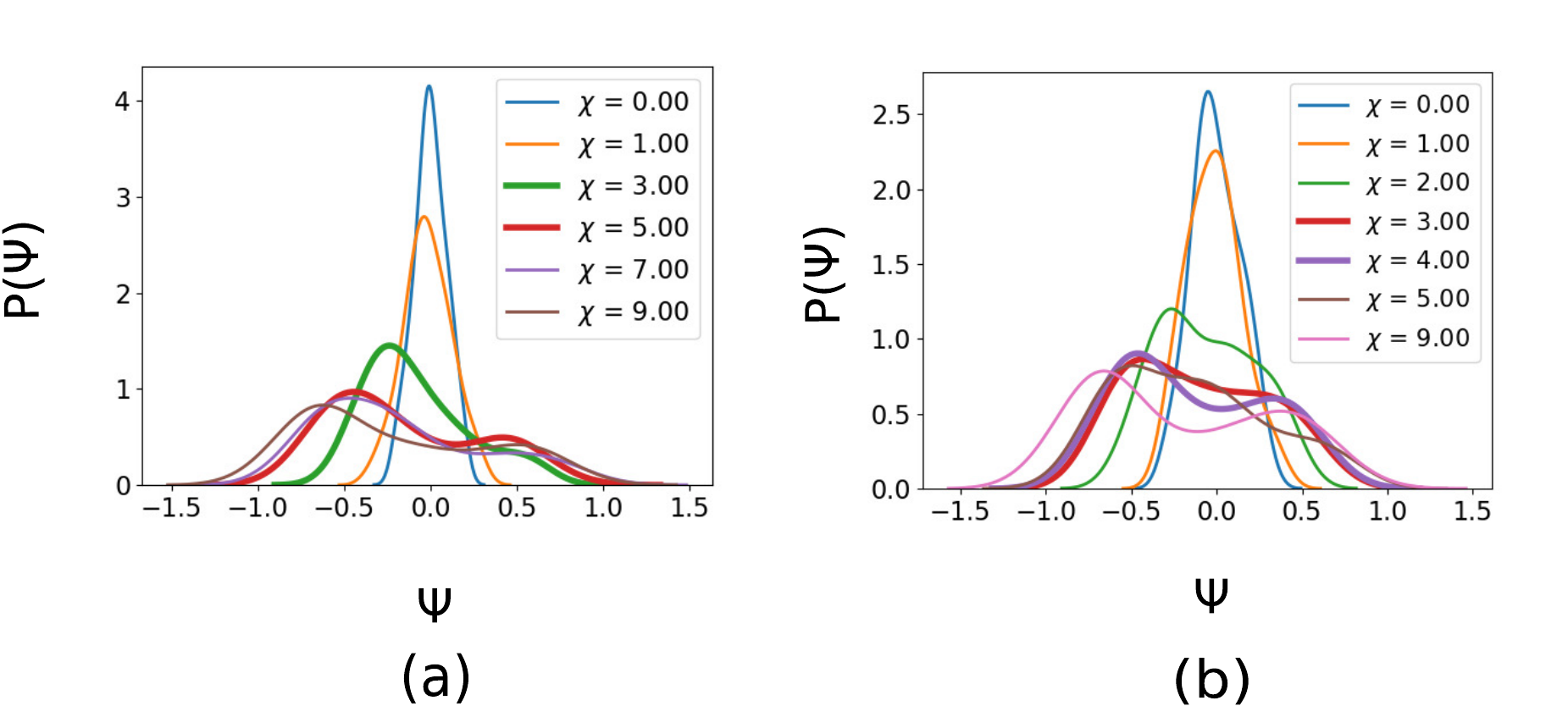}
	\caption{Distribution of $ \psi $ ($ P(\psi) $) for different aspect ratios: (a) $ L/D = 2.00 $ at $\eta = 0.35$ and (b) $ L/D = 3.00 $ at $\eta = 0.33$. Critical activity ($ \chi_{c} $) corresponds to that value of $ \chi $ from which $ P(\psi) $ develops bi-modality. Here, for $ L/D = 2.00 $, $ \chi_{c} = 3.00-5.00 $ and for $ L/D = 3.00 $, $ \chi_{c} = 2.00-3.00 $ } \label{phic}

\end{figure*}

We calculate critical activity ($ \chi_{c} $) for phase separation  by following criteria: We divide the simulation box in number of sub-cells
and calculate the quantity $ r = \frac{n_{c}-n_{h}}{n_{c}+n_{h}} $ where $n_{c}$, $n_{h}$ represent number of cold and hot particles in each sub-cell. It is then averaged over all the steady state configurations: $ \psi = \langle r \rangle _{ss}$
We then calculate distribution of $ \psi $ ($ P(\psi) $) which is unimodal for mixed system (without activity) and bi-modal for phase separated system. Critical activity ($ \chi_{c} $) 
is defined as the value of activity from which $ P(\psi) $ develops bi-modality. $ \chi_{c} $ is calculated for different aspect ratios and presented Fig. \ref{phic}. We notice some plateau regime near the peak for the  $ \chi $ just below the 
$ \chi_{c} $. This is the signature of emergence of phase separation which means exact value of $ \chi_{c} $ lies somewhere between these two lines \cite{Active-SRS}.

\begin{figure} [!htb]
	\centering
	\includegraphics[scale=0.60]{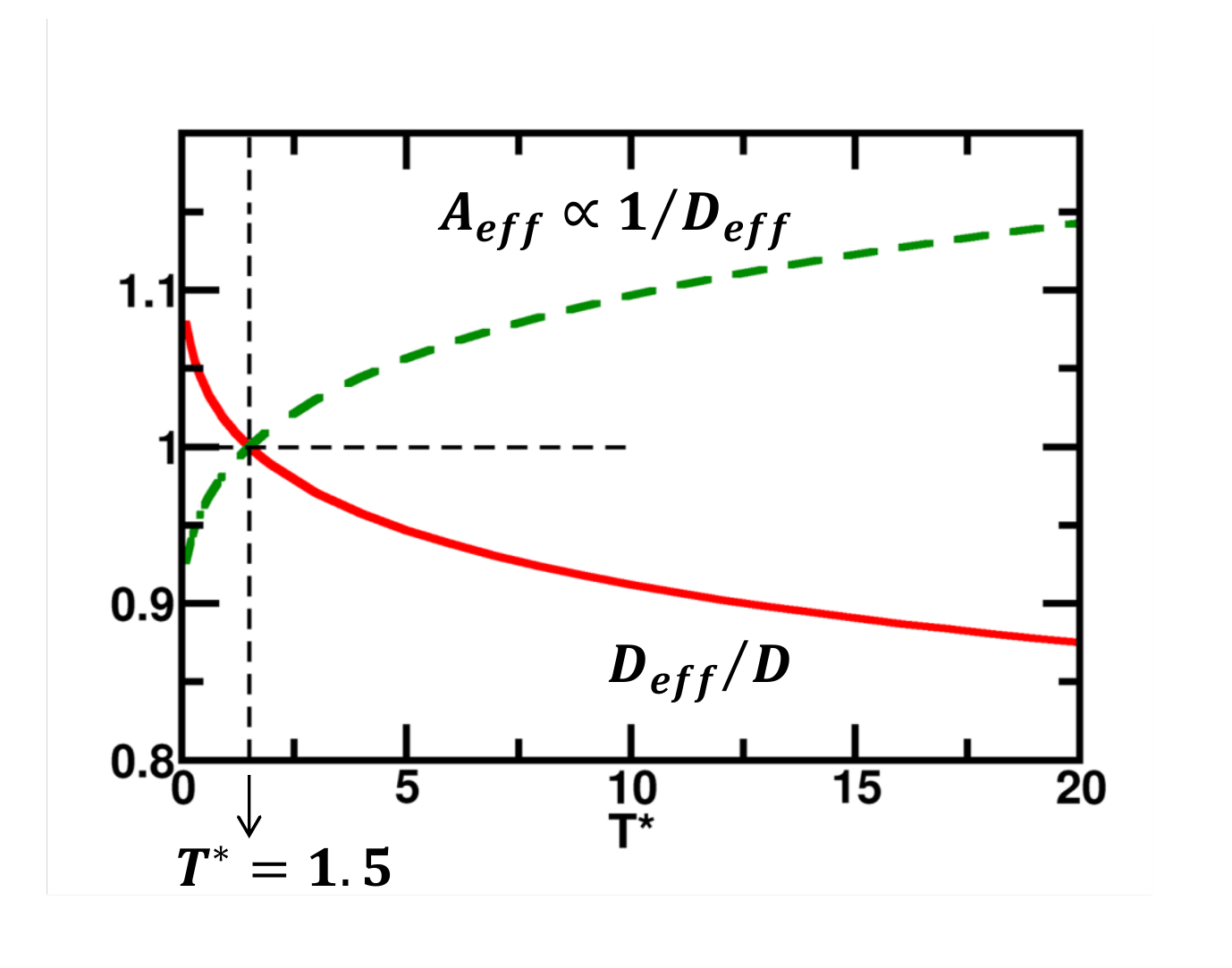}
	\caption{Temperature dependence of the effective molecular diameter $D_{eff}$ and aspect ratio $A_{eff}$ of soft repulsive spherocylinders (SRS) that can be mapped on hard spherocylinders (HSC) of aspect ratio $A_{HSC} $. We can see that $D_{eff}$ decreases (red solid line) and $A_{eff}$ (green dashed line) increases as a function of temperature irrespective of length of the spherocylinder $L$. At $T^{*} = 1.5$, $D_{eff}/D = 1 $. Therefore, at this temperature, the effective HSC fluid has the same diameter and aspect ratio as that of the SRS.} \label{d_eff}
	
\end{figure}
\begin{figure*} [!htb]
	\centering
	\includegraphics[scale=0.9]{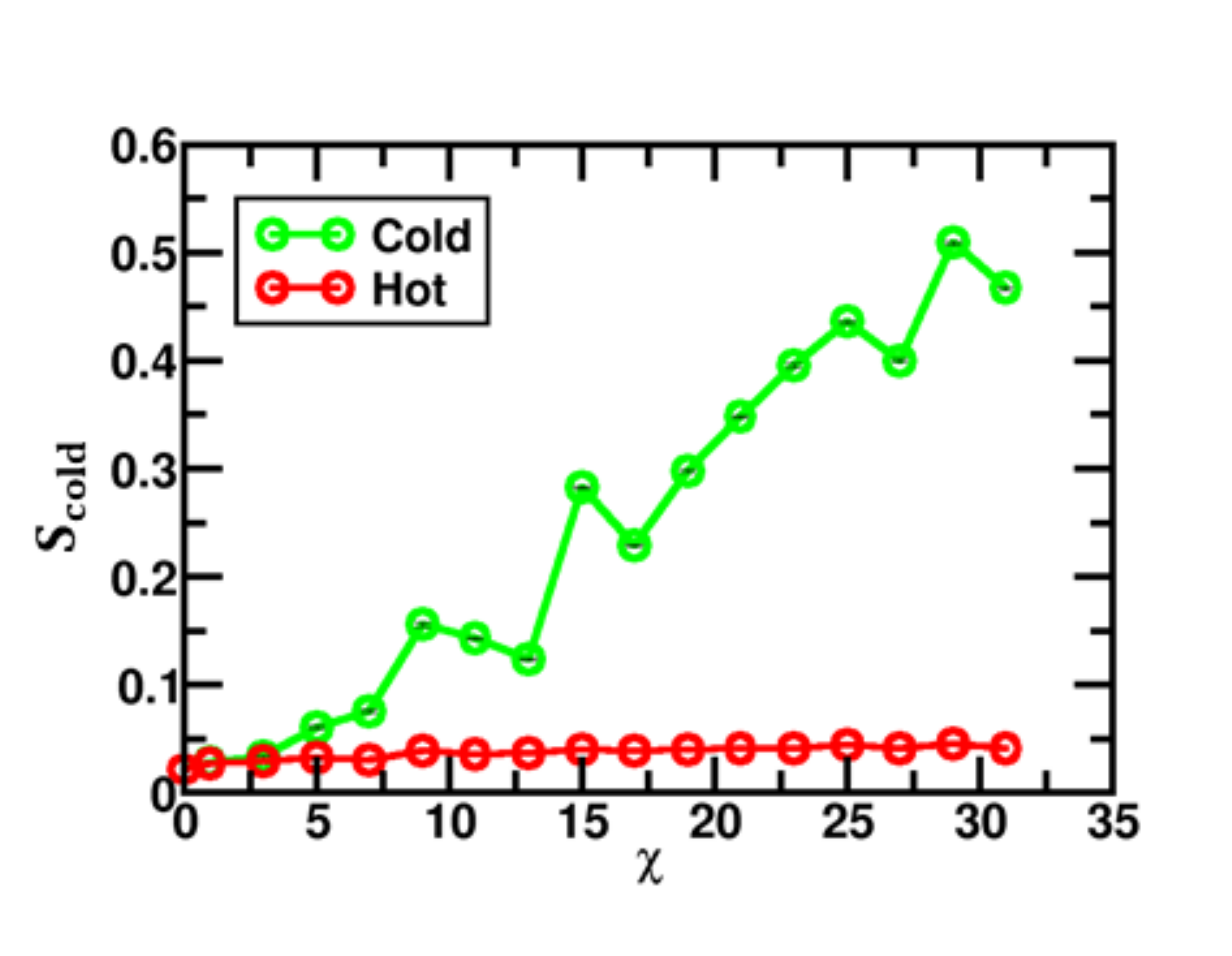}
	\caption{ Nematic order parameter for cold particles ($S_{cold}$) as a function of activity for $L/D = 2$ at $\eta = 0.35$. $S_{cold}$ starts to increase at higher activities, as mentioned in Fig. 2(b) in the main text.} \label{s-0.34-L2}

\end{figure*}

\begin{figure*} [!htb]
	\centering
	\includegraphics[scale=0.8]{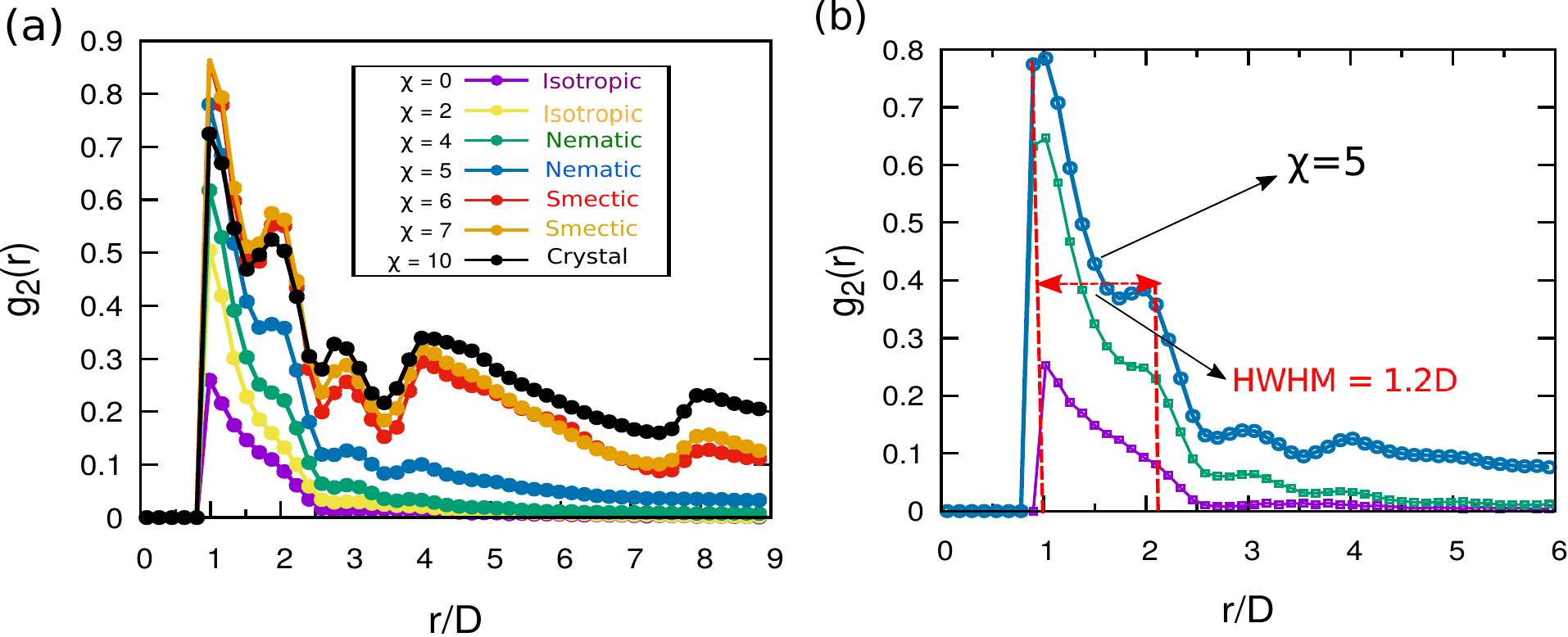}
	\caption{Orientational pair distribution function $g_{2}(r)$ for the cold particles at different activities $\chi$ for $ L/D = 3 $ at $ \eta = 0.33 $. In Fig. (a) we plot long range correlation of $g_{2}(r)$ and in Fig. (b) we show that for $\chi = 5$, the value of half width at half maxima (HWHM) is $1.2D$, indicating finite orientational correlation in the cold zone. The parameters are as the same as mentioned in Fig.3 of the main text. } \label{g-L3-b}
	
\end{figure*}

\begin{figure*} [!htb]
	\centering
	\includegraphics[scale=0.8]{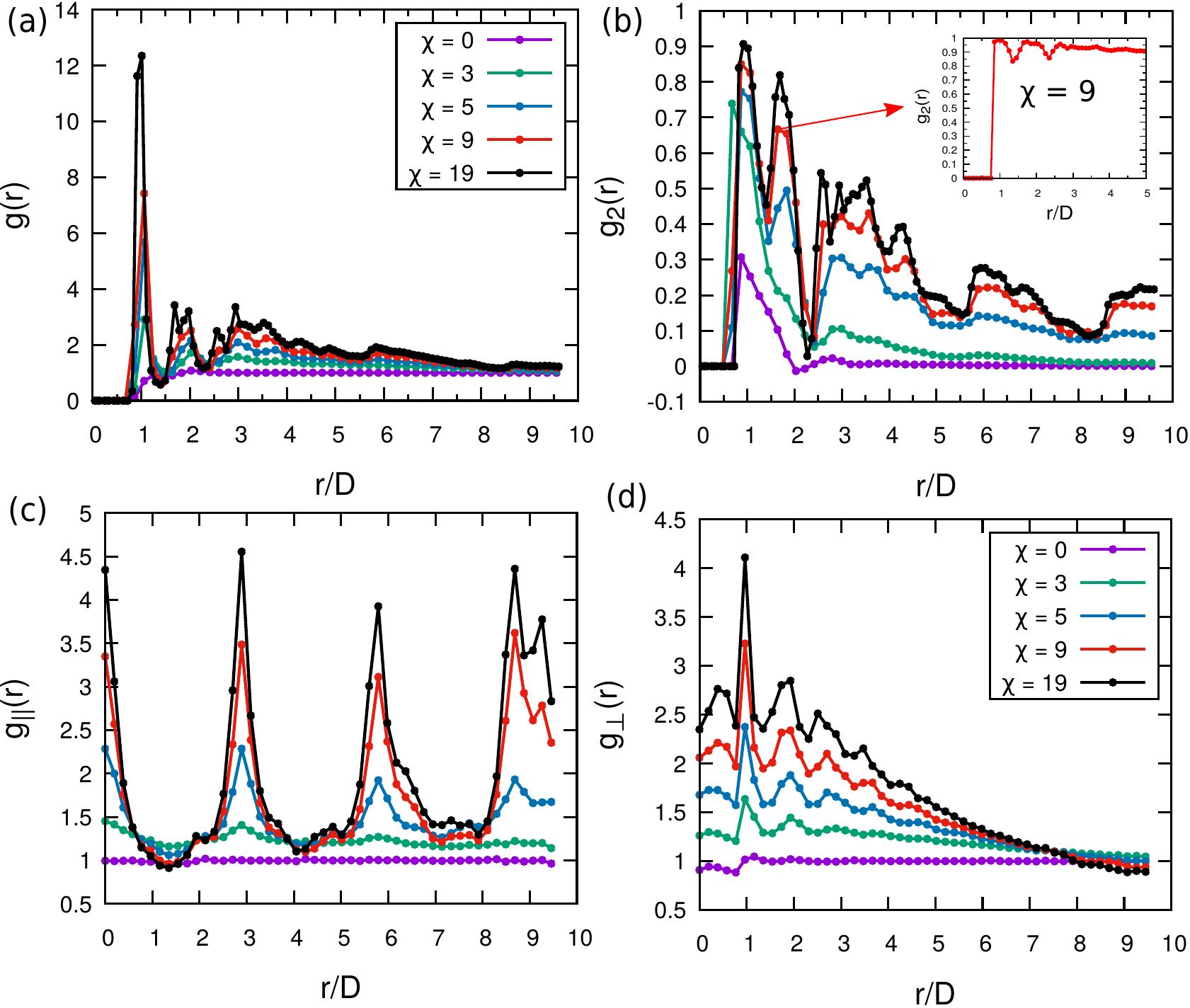}
	\caption{Pair correlation functions of the cold particles for $ L/D = 2 $ at $ \eta = 0.45 $ for several values of $\chi$. (a) The center of mass pair radial distribution function $ g(r) $, (b) orientational pair radial distribution function $ g_{2}(r) $, (c)projection of the $ g(r) $ for the distances parallel  ($ g_{\parallel} (r) $) and (d) perpendicular ($ g_{\perp} (r) $)  to the director of the spherocylinder. The inset of Fig. (b) shows $g_{2}(r)$ at $\chi = 9$ in a single cluster of a definite director. In Fig.(c) we see that, $g_{\parallel} (r)$ develops periodic oscillation from $\chi \ge 5 $, indicating the phase as smectic. The distance between two peaks in $g_{\parallel} (r)$ is $3D$ which is the end to end distance of a spherocylinder of $L/D = 2$.} \label{g-L2}
	
\end{figure*}


\begin{figure*} [!htb]
	\centering
	\includegraphics[scale=0.3]{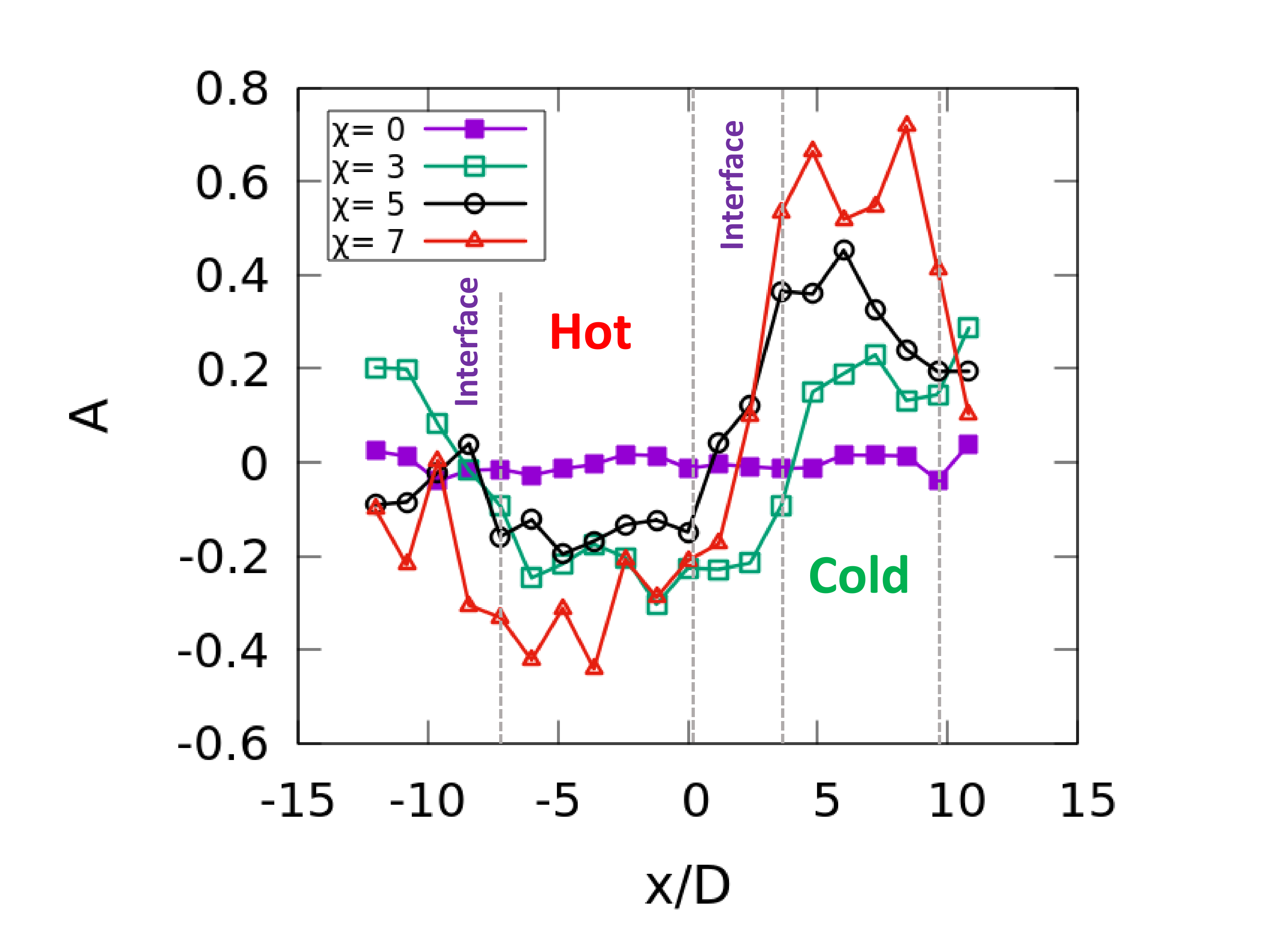}
	\caption{Pressure anisotropy $A = P_{n}-P_{t}$ for $L/D = 3$ at $\eta = 0.33$ for different values of  $\chi$ along the direction normal to the interface. The tentative location of the segregated zones are shown in the figure and the gray dashed lines indicate the zone-boundaries. We see that, pressure anisotropy acquires a positive value in the cold zone after phase separation and increases with that of activity.} \label{L3-1024-stessana}
	\end{figure*}

\begin{figure*} [!htb]
	\centering
	\includegraphics[scale=1.0]{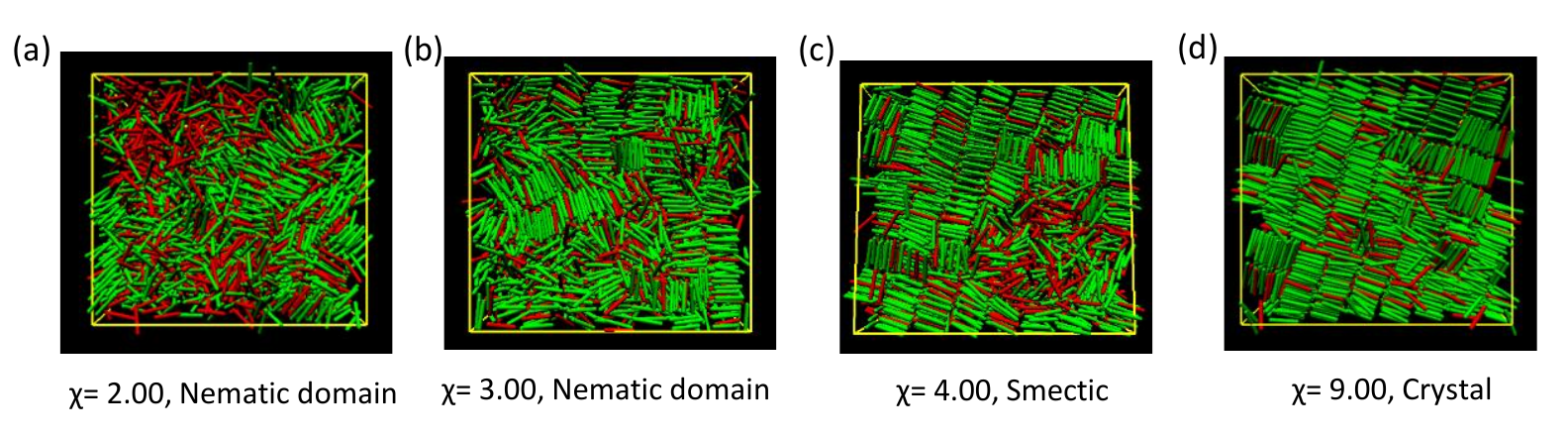}
	\caption{Snapshots of final configurations of the cold zone at different activities for $L/D = 3$ with a larger system size of $N = 3456$ at the packing fraction $\eta = 0.33$ which shows isotropic structure in equilibrium. Cold zone develops (a,b) nematic domain of different average directors (c) smectic, (d) crystal structure at the respective activities as mentioned in the legends.} \label{L3-3456-snap}
	\end{figure*}
\begin{figure*} [!htb]
	\centering
	\includegraphics[scale=0.8]{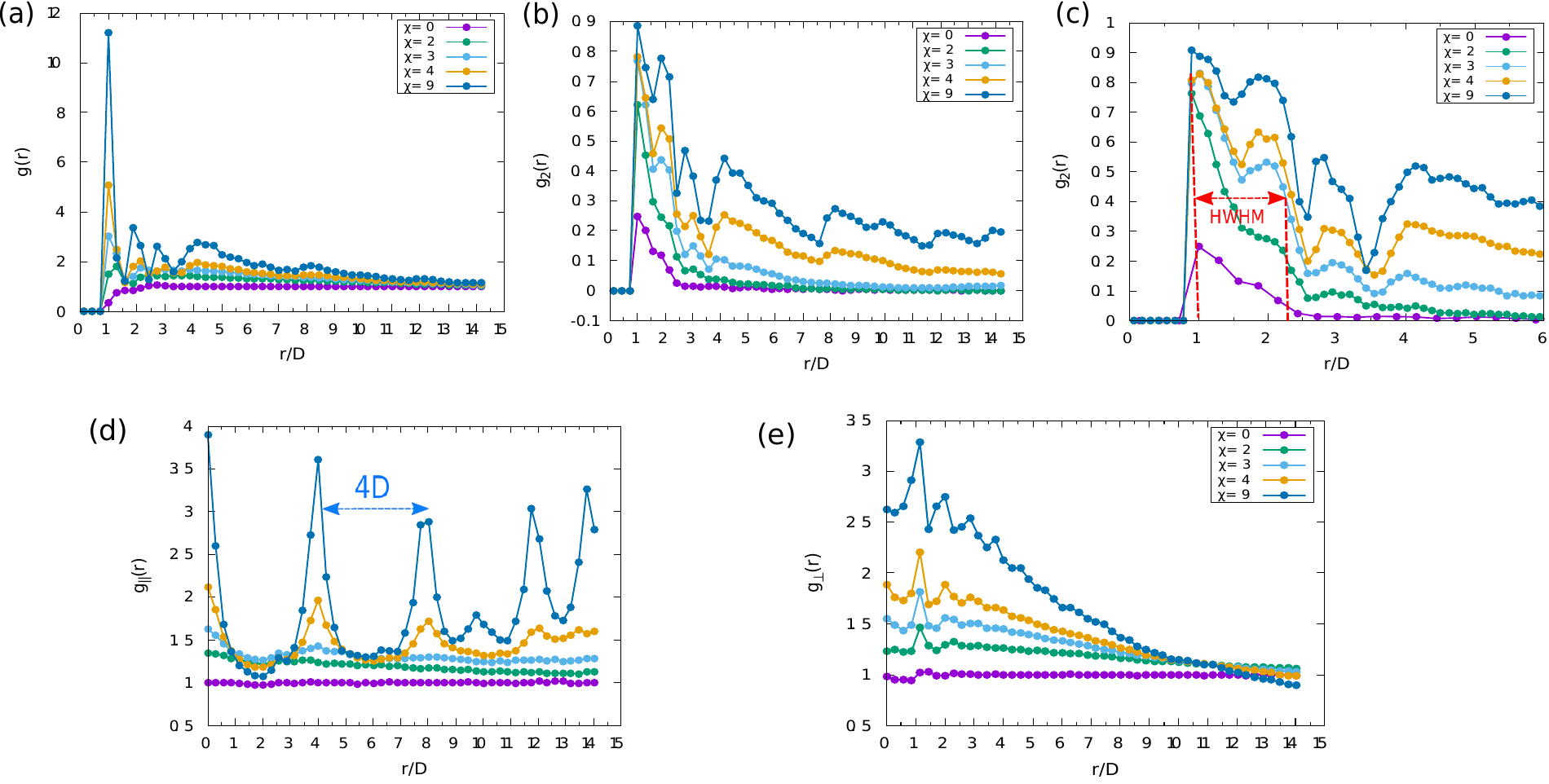}
	\caption{Pair correlation functions of the cold particles for $L/D = 3$ at larger system size with $  N = 3456$ at $ \eta = 0.33 $.(a) The center of mass pair radial distribution function $ g(r) $ and (b) orientational pair radial distribution function $ g_{2}(r) $ (d)  Projection of the $ g(r) $ for distances parallel  ($ g_{\parallel} (r) $) and (e) perpendicular ($ g_{\perp} (r) $)  to the director of spherocylinder. Figure (c) indicates orientational pair distribution function in the cold zone $ g_{2}(r) $. The red arrow indicates the range of orientational correlation in the cold cluster quantified by half width at half maxima (HWHM) which is roughly $1.3D$ for $\chi = 3$. } \label{g-L3-3456}
	
\end{figure*}


\begin{figure*} [!htb]
	\centering
	\includegraphics[scale=0.5]{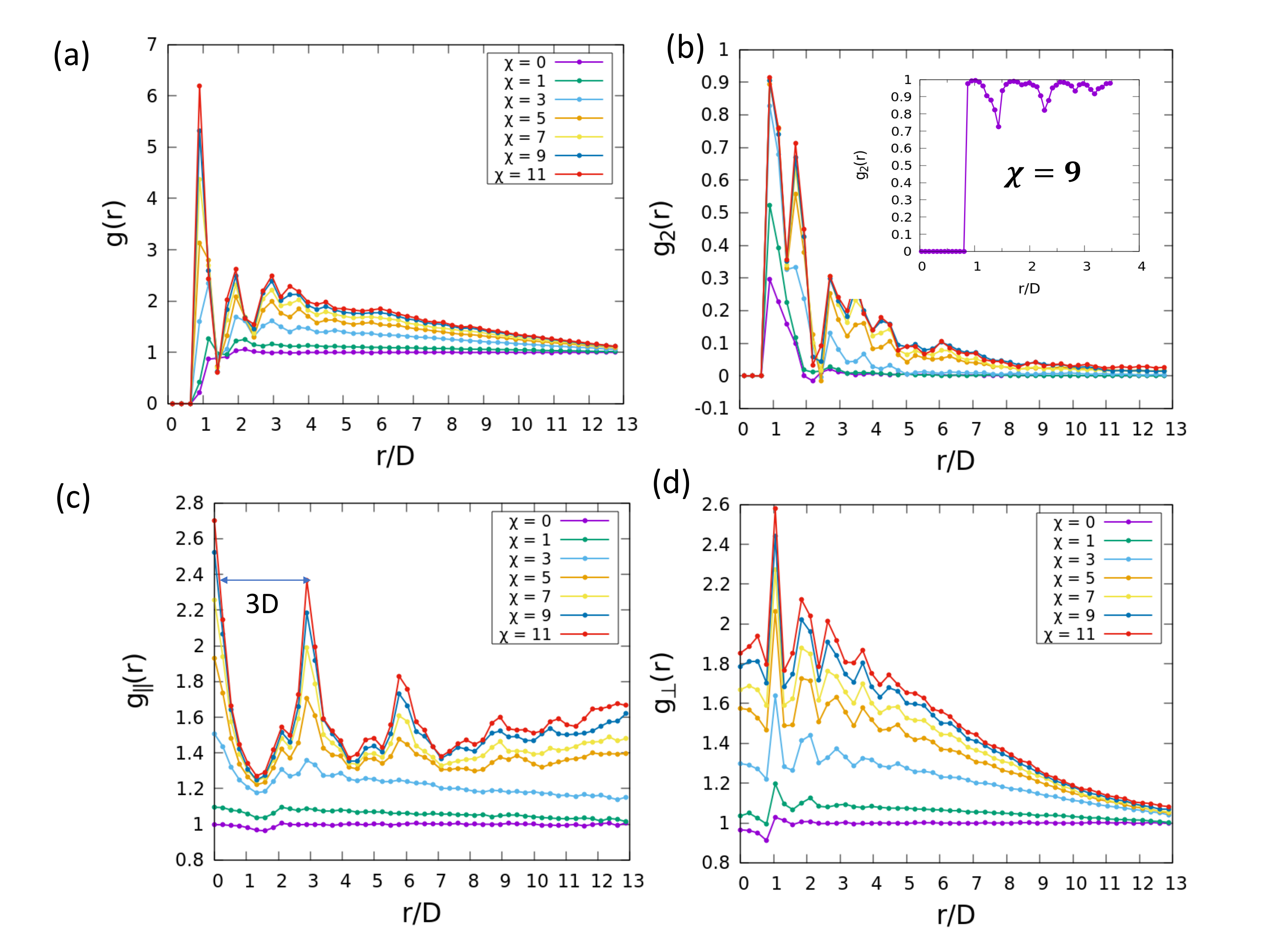}
	\caption{Pair correlation functions of the cold zone for $ L/D = 2.00 $ with $N = 4608$ at $ \eta = 0.45 $. (a) The center of mass pair radial distribution function $ g(r) $ and (b) orientational pair radial distribution function $ g_{2}(r) $ (c)  Projection of the $ g(r) $ for distances parallel  ($ g_{\parallel} (r) $) and (d) perpendicular ($ g_{\perp} (r) $)  to the director of spherocylinder.The inset of Fig. (b) shows $g_{2}(r)$ at $\chi = 9$ in a single cluster of a definite director. The periodic oscillations in $g_{\parallel} (r)$ at $\chi \ge 5 $, indicate the phase as smectic. } \label{g-L2-4608}
	
\end{figure*}


\begin{figure*} [!htb]
	\centering
	\includegraphics[scale=0.7]{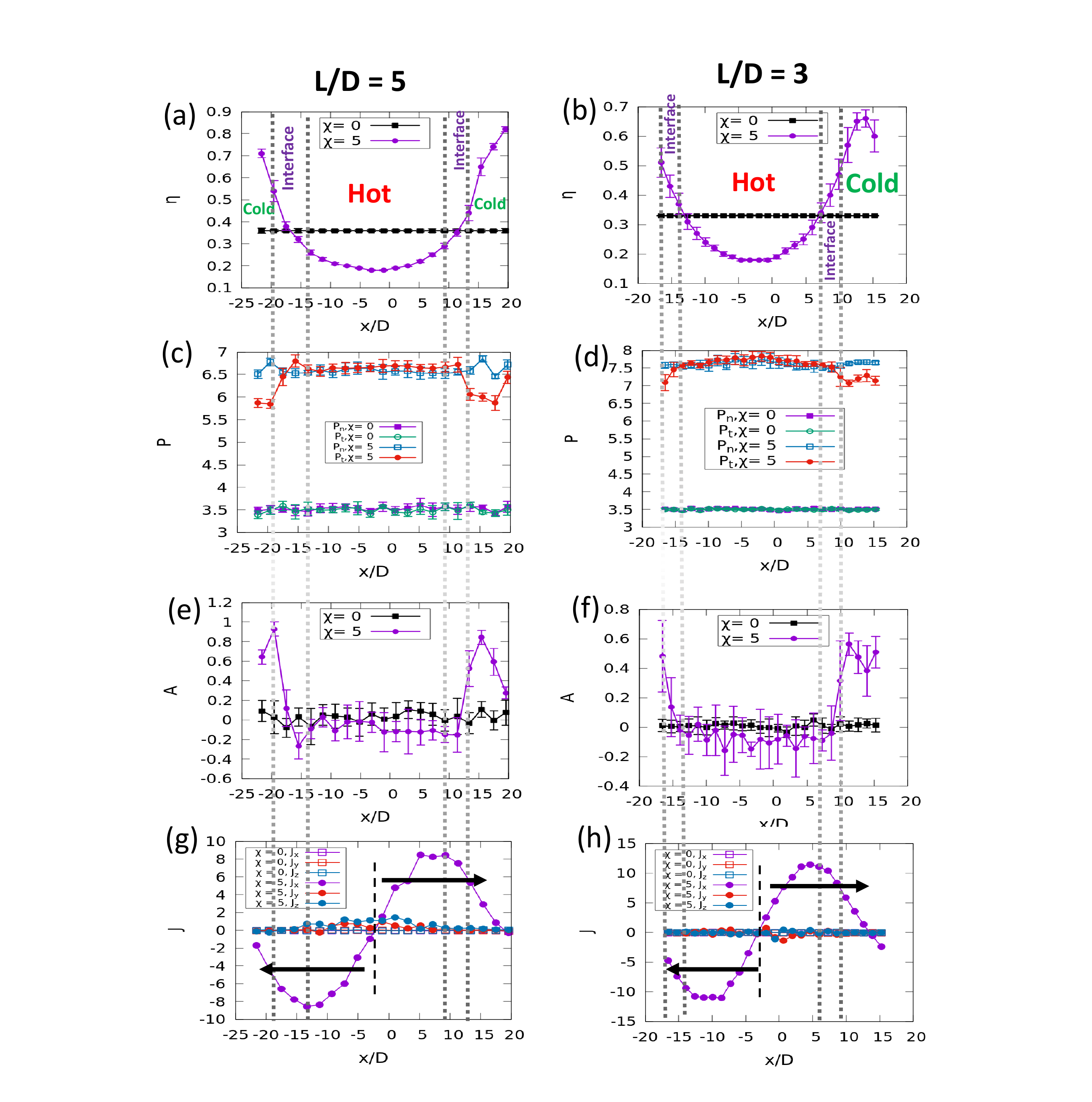}
	\caption{Stress anisotropy and local heat flux across the hot-cold interface for larger system sizes with $ N = 4096$ for $L/D = 5$ (left panel) and $N = 3456$ for $L/D = 3$ (right panel).  We plot (a, b) effective packing fraction $\eta$; (c, d) normal ($P_{n}$) and tangential ($P_{t}$) components of the stress tensor; (e, f) pressure anisotropy $A = P_{n}-P_{t}$ and (g, h) local heat flux $J$ along the direction perpendicular to the interfacial plane. 
	The grey dotted lines indicate boundaries of different zones. 
	The black dotted lines indicate the location where heat flux becomes $0$ . The arrows indicate the directions of the heat flow depending on the sign of $J$.
    Here also we find a positive pressure anisotropy at the interface that extends in the cold zone 
 	and a finite heat flux flowing from the bulk hot zone to the bulk cold zone as we have seen earlier in Fig. 4 of the main text.}
	\label{stress-ana-bigsys}
	\end{figure*}


\end{document}